\documentstyle[11pt,aaspp]{article}
\slugcomment{Caltech GRP-375; MIT-AT-94-01; IASSNS-AST-94/1}
\begin{document}
\title{Cosmological Perturbation Theory in the Synchronous vs. Conformal
    Newtonian Gauge}
\author{Chung-Pei Ma}
\affil{Theoretical Astrophysics 130-33, California Institute of
Technology, Pasadena, CA 91125}
\centerline{and}
\author{Edmund Bertschinger\altaffilmark{1}}
\affil{Department of Physics, Massachusetts Institute of Technology,
Cambridge, MA 02139}
\altaffiltext{1}{Also Institute for Advanced Study, Princeton, NJ 08540}
\begin{abstract}
We present a systematic treatment of the linear theory of scalar
gravitational perturbations in the synchronous gauge and the conformal
Newtonian (or longitudinal) gauge.  We first derive the transformation
law relating the two gauges.  We then write down in parallel in both
gauges the coupled, linearized Boltzmann, Einstein and fluid equations
that govern the evolution of the metric perturbations and the density
fluctuations of the particle species.  The particle species considered
include cold dark matter (CDM), baryons, photons, massless neutrinos,
and massive neutrinos (a hot dark matter or HDM candidate), where the
CDM and baryon components are treated as fluids while a detailed
phase-space description is given to the photons and neutrinos.  The
linear evolution equations presented are applicable to any $\Omega=1$ model
with CDM or a mixture of CDM and HDM.  Isentropic initial conditions
on super-horizon scales are derived.  The equations are solved
numerically in both gauges for a CDM+HDM model with $\Omega_{\rm
cold}=0.65,$ $\Omega_{\rm hot}=0.3$, and $\Omega_{\rm baryon}=0.05$.
We discuss the evolution of the metric and the density perturbations
and compare their different behaviors outside the horizon in the two
gauges.  In a companion paper we integrate the geodesic equations for
the neutrino particles in the perturbed conformal Newtonian background
metric computed here.  The purpose is to obtain an accurate sampling
of the neutrino phase space for the HDM initial conditions in $N$-body
simulations of the CDM+HDM models.
\end{abstract}
\centerline{Submitted to {\it The Astrophysical Journal}}
\centerline{January 1994}

\section{Introduction}
The theory of galaxy formation based on gravitational instability is
aimed at describing how primordially-generated fluctuations in matter
and radiation grow into galaxies and clusters of galaxies due to
self-gravity.  A perturbation theory can be formulated when the
amplitudes of the fluctuations are small, and the growth of the
fluctuations can be solved from the linear theory.  Such linear theory
for a perturbed Friedmann-Robertson-Walker universe was first
developed by Lifshitz (1946), later reviewed in Lifshitz \&
Khalatnikov (1963).  The subsequent work can be found summarized in
the textbooks by Weinberg (1972) and Peebles (1980), and in the
reviews by Kodama \& Sasaki (1984) and Mukhanov, Feldman \&
Brandenberger (1992).

In the early universe, gravitational perturbations are inflated to
wavelengths beyond the horizon at the end of the inflationary epoch.
Fluctuations of a given length scale reenter the horizon at a later
time when the horizon has grown to the size of the fluctuations.
Although the process of galaxy formation in recent epochs is well
described by Newtonian gravity (and other microphysical processes such
as hydrodynamics), a general relativistic treatment is required for
perturbations on scales larger than the horizon size before the
horizon crossing time.  The use of general relativity brought in the
issue of gauge freedom which has caused some confusion over the years.
Lifshitz (1946) adopted the ``synchronous gauge'' for his coordinate
system, which has since become the most commonly used gauge for
cosmological perturbation theories.  However, some complications
associated with this gauge such as the appearance of coordinate
singularities and spurious gauge modes prompted Bardeen (1980) and
others (e.g. Kodama \& Sasaki, 1984) to formulate alternative
approaches that deal only with gauge-invariant quantities.  A thorough
review of the gauge-invariant perturbation theory and its appliaiton
to texture-seeded structure formation models is given by Durrer (1993).
Another possibility is to adopt a different gauge.  We will discus in
detail in this paper the conformal Newtonian (or the longitudinal) gauge
(Mukhanov et al. 1992), which is a particularly convenient gauge to use
for scalar perturbations.

This paper serves two purposes.  First, it is an independent paper in
which we present and compare a systematic treatment of the linear theory
of scalar isentropic gravitational perturbations in the synchronous and
conformal Newtonian gauges.  The coupled, linearized Einstein, Boltzmann,
and fluid equations for the metric and density perturbations are presented
in parallel in the two gauges.  We give a full discussion of the evolution
of five particle species: cold dark matter (CDM), hot dark matter (HDM,
i.e., massive neutrinos), baryons, photons, and massless neutrinos.
The CDM and the baryon components behave like collisionless and
collisional fluids, respectively, while the photons and the neutrinos
require a phase-space description governed by the Boltzmann transport
equation.  We also derive analytically the time dependence of the
perturbations on scales larger than the horizon to illustrate the
dependence on the gauge choice.  This information is needed in the
initial conditions for the numerical integration of the evolution
equations.  The linear theory discussed in this paper can be applied
to $\Omega=1$ models with various dark matter compositions, e.g., pure
CDM, pure HDM, or a mixture of CDM and HDM.

This paper also serves as a companion paper to Ma \& Bertschinger
(1994) in which we reported the main results from our linear
calculation of the full neutrino phase space in a CDM+HDM model with
$\Omega_{\rm cold}=0.65$, $\Omega_{\rm hot}=0.3$, $\Omega_{\rm
baryon}=0.05$, and $H_0=50$ km s$^{-1}$ Mpc$^{-1}$.  (The
corresponding neutrino mass is $m_\nu = 6.985$ eV.)  The motivation
was to obtain an accurate sampling of the neutrino phase space for the
HDM initial conditions in $N$-body simulations of structure formation
in CDM+HDM models.  We adopted a two-step Monte Carlo procedure to
achieve this goal: (1) Integrate the coupled, linearized Boltzmann,
Einstein, and fluid equations for all particle species in the model
(i.e., CDM, HDM, photons, baryons and massless neutrinos) to obtain
the evolution of the metric perturbations; (2) Follow the trajectories
of individual neutrinos by integrating the geodesic equations using
the metric computed in (1).  Since no coordinate singularities occur
in the conformal Newtonian gauge and the geodesic equations have
simple forms, the geodesic integration in step (2) was carried out in
this gauge, starting shortly after neutrino decoupling at redshift
$z\sim 10^9$ until $z = 13.5$.  We focus on step (1) in this paper.
Following historical precedents, we first developed the code for the
Boltzmann integration in the synchronous gauge.  The transformation
relating the synchronous gauge and the conformal Newtonian gauge was
then derived and used to compute the metric perturbations in the
latter gauge for step (2) of the calculation.
Subsequently we developed a code to perform the full integration in
the conformal Newtonian gauge.

The organization of this paper is as follows.  In \S 2 we write down
the metric for the two gauges and summarize their properties.  In \S
3, we derive the gauge transformation relating two arbitrary gauges
and obtain the transformation between the synchronous and the
conformal Newtonian gauges.  The linearized evolution equations for
the metric and the density perturbations are given in \S\S 4 and 5.
Section 4 discusses the Einstein equations with emphasis on the source
terms, the energy-momentum tensor, in the two gauges.  The perturbed
fluid equations are derived from the energy-momentum conservation,
which are applied to CDM and the baryons in \S 5.  The rest of \S 5
contains detailed treatments of the photon and neutrino phase space
distributions, recombination, and the coupling of photons and baryons.
The photon and neutrino distribution functions are expanded in
Legendre polynomials, reducing the linearized Boltzmann equation to a
set of coupled ordinary differential equations for the expansion
modes.  The massive neutrinos require a slightly more complicated
treatment due to the nontrivial time dependence of the energy-momentum
relation.  Section 6 discusses the behavior of the perturbations
before horizon crossing.  The necessary initial conditions for the
variables in the two gauges are given.  Section 7 presents the
numerical results for the evolution of the perturbations in our
CDM+HDM model.

\section{The Two Gauges}
We consider only spatially flat ($\Omega=1$) background spacetimes
with isentropic scalar metric perturbations.  The spacetime
coordinates are denoted by $x^\mu$, $\mu\in(0,1,2,3)$, where $x^0$ is
the time component and $x^i$, $i\in(1,2,3)$ are the spatial components
in Cartesian coordinates.  Greek letters $\alpha,\beta,\gamma$ and so
on always run from 0 to 3, labeling the four spacetime-coordinates;
Roman letters such as $i, j, k$ always run from 1 to 3, labeling the
spatial parts of a four-vector.  Repeated indices are summed.  Since
our interests lie in the physics in an expanding universe, we use
comoving coordinates $x^\mu = (\tau,\vec{x})$ with the expansion
factor $a(\tau)$ of the universe factored out.  The comoving
coordinates are related to the proper time and positions $t$ and
$\vec{r}$ by $dx^0 =d\tau = dt/a(\tau)$, $d\vec{x}=d\vec{r}/a(\tau)$.
Dots will denote derivatives with respect to $\tau$: $\dot
a\equiv\partial a/\partial\tau$.  The speed of light $c$ is set to
unity.

The components $g_{00}$ and $g_{0i}$ of the metric tensor in the
synchronous gauge is by definition unperturbed.  The line element is
given by
\begin{equation}
  ds^2 = a^2(\tau)\{-d\tau^2 + (\delta_{ij} + h_{ij})dx^i dx^j\}\,.
\end{equation}
The metric perturbation $h_{ij}$ can be decomposed into a trace part
$h \equiv h_{ii}$ and a traceless part consisting of three
pieces, $h^\parallel_{ij}, h^\perp_{ij}$, and $h^T_{ij}$, where
$h_{ij}=h\delta_{ij}/3 + h^\parallel_{ij}+h^\perp_{ij}+h^T_{ij}$.
By definition, the divergences of $h^\parallel_{ij}$ and
$h^\perp_{ij}$ (which are vectors) are longitudinal and transverse,
respectively, and $h^T_{ij}$ is transverse, satisfying
\begin{equation}
\label{decomp}
     \epsilon_{ijk} \partial_j \partial_l h^\parallel_{lk} = 0\,,
	\qquad \partial_i\partial_j h^\perp_{ij} = 0\,,\qquad
     \partial_i h^T_{ij} = 0 \,.
\end{equation}
It then follows that $h^\parallel_{ij}$ can be written in terms
of some scalar field $\Lambda$ and $h^\perp_{ij}$ in terms of
some divergenceless vector $\vec{A}$ as
\begin{eqnarray}
        h^\parallel_{ij} &=& \left( \partial_i\partial_j -
	{1\over 3} \delta_{ij} \nabla^2 \right) \Lambda\,, \nonumber\\
        h^\perp_{ij} &=& \partial_i A_j + \partial_j A_i\,,\qquad
        \partial_i A_i = 0\,.
\end{eqnarray}
The two scalar fields $h$ and $\Lambda$ (or $h^\parallel_{ij}$)
characterize the scalar mode of the metric perturbations,
while $A_i$ (or $h^\perp_{ij}$) and $h^T_{ij}$ represent the vector
and the tensor modes, respectively.

We will be working in the Fourier space $k$ in this paper.
We introduce two fields $h(\vec{k},\tau)$ and $\eta(\vec{k},\tau)$ in
$k$-space and write the scalar mode of $h_{ij}$ as a Fourier integral
\begin{equation}
\label{hijk}
	h_{ij}(\vec{x},\tau) = \int d^3k e^{i\vec{k}\cdot\vec{x}}
	\left\{ \hat{k}_i\hat{k}_j h(\vec{k},\tau) +
	(\hat{k}_i\hat{k}_j - {1 \over 3}\delta_{ij})\,
        6\eta(\vec{k},\tau) \right\} \,,\quad \vec{k} = k\hat{k} \,.
\end{equation}
Note that $h$ is used to denote the trace of $h_{ij}$ in both the
real space and the Fourier space.

In spite of its wide-spread use, there are serious disadvantages
associated with the synchronous gauge.  Since the choice of the
initial hypersurface and its coordinate assignments are arbitrary,
the synchronous gauge conditions do not fix the gauge degrees of
freedom completely.  Such residual gauge freedom is manifested in the
spurious gauge modes contained in the solutions to the equations
for the density perturbations.  The appearance of these modes has
caused some confusion over the years and prompted Bardeen (1980)
to formulate alternative
approaches that deal only with gauge-invariant quantities.  Another
difficulty with the synchronous gauge is that since the coordinates
are defined by freely falling observers, coordinate singularities
arise when two observers' trajectories intersect each other: a point
in spacetime will have two coordinate labels.  A different initial
hypersurface of constant time has to be chosen to remove these
singularities.

The conformal Newtonian gauge (also known as the longitudinal gauge)
advocated by Mukhanov et al. (1992) is a particularly simple gauge to use
for the scalar mode of metric perturbations.  The perturbations are
characterized by two scalar potentials $\psi$ and $\phi$ which appear
in the line element as
\begin{equation}
\label{conformal}
    ds^2 = a^2(\tau)\left\{ -(1+2\psi)d\tau^2 +
        (1-2\phi)dx^i dx_i \right\} \,.
\end{equation}
It should be emphasized that the conformal Newtonian gauge is a
restricted gauge since the metric is applicable only for the scalar
mode of the metric perturbations; the vector and the tensor degrees of
freedom are eliminated from the beginning.  Nonetheless, it can be
easily generalized to include the vector and the tensor modes.
We will confine our discussion here to the scalar perturbations only.

One advantage of working in this gauge is that the metric tensor
$g_{\mu\nu}$ is diagonal.  This simplifies the calculations and leads
to simple geodesic equations (Ma \& Bertschinger 1994).  Another
advantage is that $\psi$ plays the role of the gravitational potential
in the Newtonian limit and thus has a simple physical interpretation.
Moreover, the two scalar potentials $\psi$ and $\phi$ in this gauge
are identical (up to a minus sign) to the gauge-invariant variables
$\Phi_A$ and $\Phi_H$ Bardeen constructed (1980).  No gauge modes are
present in this gauge to obscure the meaning of the physical modes
since the gauge freedom is completely fixed for $\Omega=1$ aside from
the addition of spatial constants to $\psi$ and $\phi$.  The second
scalar $\phi$ is required when the energy-momentum tensor
$T^\mu{}_{\!\nu}$ contains a nonvanishing traceless and longitudinal
component.  As we will see in equation (\ref{theta}) and the Einstein
equation (\ref{ein-cond}), this component provides the source term for
the constraint equation for $(\phi-\psi)$.  When this component is
absent, the two scalar potentials $\psi$ and $\phi$ are identical.

\section{Gauge Transformations}
In this section we first derive the transformation law relating two
arbitrary gauges.  From it, the gauge transformation relating the
synchronous gauge and the conformal Newtonian gauge is readily obtained.

A perturbed flat Friedmann-Robertson-Walker metric can be written in
general as
\begin{eqnarray}
\label{perturb}
  g_{00} &=& -a^2(\tau)\,\left\{ 1+2\psi(\vec{x},\tau) \right\}\,,
	\nonumber\\
  g_{0i} &=& -a^2(\tau)\,w_i(\vec{x},\tau)\,,\\
  g_{ij} &=& a^2(\tau)\,\left\{ [1-2\phi(\vec{x},\tau)] \delta_{ij}
        + \sigma_{ij}(\vec{x},\tau) \right\}\,,\qquad \sigma_{ii}=0
	\nonumber
\end{eqnarray}
where the functions $\psi, \phi, w_i$ and $\sigma_{ij}$ represent
metric perturbations about the Robertson-Walker spacetime
and are assumed to be small compared with unity.  The trace part of
the perturbation to $g_{ij}$ is absorbed in $\phi$, and $\sigma_{ij}$
is taken to be traceless.

Consider a general coordinate transformation from a coordinate
system $x^\mu$ to another $\hat{x}^\mu$
\begin{equation}
        x^\mu \rightarrow \hat{x}^\mu = x^\mu + d^\mu(x^\nu)\,.
\end{equation}
We write the time and the spatial parts separately as
\begin{eqnarray}
\label{shift}
        \hat{x}^0 &=& x^0 + \alpha(\vec{x},\tau)\,, \nonumber\\
        \hat{\vec{x}} &=& \vec{x} + \vec{\nabla}\beta(\vec{x},\tau)
                + \vec{\epsilon}\,(\vec{x},\tau)\,,\quad
                \vec{\nabla}\cdot\vec{\epsilon} = 0\,,
\end{eqnarray}
where the vector $\vec{d}$ has been decomposed into a longitudinal
component $\vec{\nabla}\beta\,$
($\vec{\nabla}\times\vec{\nabla}\beta=0$)
and a transverse component
$\vec{\epsilon}\,$ ($\vec{\nabla}\cdot\vec{\epsilon}=0$).  The
requirement that $ds^2$ be invariant under this coordinate
transformation leads to
\begin{equation}
        \hat{g}_{\mu\nu}(x) = g_{\mu\nu}(x)
        - g_{\mu\beta}(x) \partial_\nu d^\beta
        - g_{\alpha\nu}(x) \partial_\mu d^\alpha
        - d^\alpha \partial_\alpha g_{\mu\nu}(x) + O(d^2) \,.
\end{equation}
We note that both sides of this equation are evaluated at the same
coordinate values $x$ in the two gauges, which do not correspond to
the same physical point in general.  Assuming $d^\mu$ to be of the
same order as the metric perturbations $\psi,w_i,\phi$ and
$\sigma_{ij}$, the metric perturbations in the two coordinate systems
are related to first order in the perturbed quantities by
\begin{mathletters}
\begin{eqnarray}
    \hat{\psi}(\vec{x},\tau) &=&
       \psi (\vec{x},\tau) - \dot{\alpha}(\vec{x},\tau)
       - {\dot{a}\over a} \alpha(\vec{x},\tau)\,,
	\label{trans1a} \\
    \hat{w}_i(\vec{x},\tau) &=&
          w_i(\vec{x},\tau) + \partial_i \alpha(\vec{x},\tau)
          - \partial_i \dot{\beta}(\vec{x},\tau)
          - \dot{\epsilon}_i(\vec{x},\tau)\,,
	\label{trans1b} \\
    \hat{\phi}(\vec{x},\tau) &=&
          \phi (\vec{x},\tau) + {1 \over 3}\nabla^2\beta(\vec{x},\tau)
          + {\dot{a} \over a} \alpha(\vec{x},\tau)\,,
	\label{trans1c} \\
    \hat{\sigma}_{ij}(\vec{x},\tau) &=&
          \sigma_{ij}(\vec{x},\tau) - 2\left\{ \left(
        \partial_i\partial_j - {1\over 3}\delta_{ij}\nabla^2 \right)
        \beta(\vec{x},\tau) + {1\over 2}\left( \partial_i\epsilon_j +
        \partial_j\epsilon_i \right) \right\}\,. \label{trans1d}
\end{eqnarray}
\label{trans1}
\end{mathletters}
We can further decompose the transformations of $w_i$ and $\sigma_{ij}$
above into longitudinal and transverse parts:
\begin{eqnarray}
\label{w}
    \hat{w}^\parallel_i(\vec{x},\tau) &=&
          w^\parallel_i(\vec{x},\tau) + \partial_i \alpha(\vec{x},\tau)
          - \partial_i \dot{\beta}(\vec{x},\tau)\,, \nonumber\\
    \hat{w}^\perp_i(\vec{x},\tau) &=&
          w^\perp_i(\vec{x},\tau) - \dot{\epsilon}_i(\vec{x},\tau)\,,
\end{eqnarray}
and
\begin{eqnarray}
\label{sigma}
    \hat{\sigma}^\parallel_{ij}(\vec{x},\tau) &=&
        \sigma^\parallel_{ij}(\vec{x},\tau) - 2\left(\partial_i\partial_j
        - {1\over 3}\delta_{ij}\nabla^2 \right) \beta(\vec{x},\tau)\,,
        \nonumber\\
    \hat{\sigma}^\perp_{ij}(\vec{x},\tau) &=&
          \sigma^\perp_{ij}(\vec{x},\tau) -
          (\partial_i\epsilon_j + \partial_j\epsilon_i)\,, \nonumber\\
    \hat{\sigma}^T_{ij}(\vec{x},\tau) &=& \sigma^T_{ij}(\vec{x},\tau) \,,
\end{eqnarray}
where $w_i = w^\parallel_i + w^\perp_i$, $\sigma_{ij} =
\sigma^\parallel_{ij}+\sigma^\perp_{ij}+\sigma^T_{ij}\,$,
and $\sigma^\parallel_{ij}\,$, $\sigma^\perp_{ij}$ and $\sigma^T_{ij}$
obey equation (\ref{decomp}).  Equations (\ref{trans1})--(\ref{sigma})
describe the transformation of metric perturbations under a general
infinitesimal coordinate transformation.

We can now use equations (\ref{trans1}) to relate the scalar metric
perturbations $(\phi, \psi)$ in the conformal Newtonian gauge to $h_{ij}
=h\delta_{ij}/3+h^\parallel_{ij}$ in the synchronous gauge.  Let $
\hat{x}^\mu$ denote the synchronous coordinates and $x^\mu$ the
conformal Newtonian coordinates with $\hat{x}^\mu=x^\mu+d^\mu$.  From
equations (\ref{w}) and (\ref{sigma}), we find
\begin{mathletters}
\begin{eqnarray}
       \alpha(\vec{x},\tau) &=& \dot{\beta}(\vec{x},\tau) + \chi(\tau)
		 \label{consyn1a} \,,\label{alpha}\\
       \epsilon_i(\vec{x},\tau) &=& \epsilon_i(\vec{x})
		 \label{consyn1b} \,,\\
       h^\parallel_{ij}(\vec{x},\tau) &=& -2\left(
            \partial_i\partial_j - {1\over 3}\delta_{ij}\nabla^2 \right)
            \beta(\vec{x},\tau)  \label{consyn1c} \,,\\
       \partial_i\epsilon_j + \partial_j\epsilon_i &=& 0\,,
		\label{consyn1d}
\end{eqnarray}
\end{mathletters}
where $\chi(\tau)$ is an arbitrary function of time, reflecting the
gauge freedom associated with the coordinate transformation:
$\hat{x}^0 = x^0 + \chi(\tau)$, $\hat{x}^i = x^i$.
This transformation corresponds to a global redefinition of the units
of time with no physical significance; therefore we shall set $\chi=0$
from now on.  From equations (\ref{trans1a}) and (\ref{trans1c}) we then
obtain
\begin{eqnarray}
\label{consyn2}
       \psi(\vec{x},\tau) &=&
            +\ddot{\beta}(\vec{x},\tau) +
            {\dot{a}\over a}\dot{\beta}(\vec{x},\tau)\,,\nonumber\\
       \phi(\vec{x},\tau) &=& -{1\over 6} h(\vec{x},\tau) -
	    {1 \over 3}\nabla^2\beta(\vec{x},\tau)
	    - {\dot{a}\over a} \dot{\beta}(\vec{x},\tau)\,,
\end{eqnarray}
and $\beta$ is determined by $h^\parallel$ in equation (\ref{consyn1c}).

In terms of $h$ and $\eta$ introduced in equation (\ref{hijk}),
$h^\parallel_{ij}$ in the synchronous gauge is given by
\begin{equation}
\label{hpara}
        h^\parallel_{ij}(\vec{x},\tau) = \int d^3k
           \,e^{i\vec{k}\cdot\vec{x}}\,(\hat{k}_i\hat{k}_j -
	   {1 \over 3}\delta_{ij})
	   \left\{ h(\vec{k},\tau) + 6\eta(\vec{k},\tau) \right\}\,,
	   \quad \vec{k} = k\hat{k} \,.
\end{equation}
Comparing $h^\parallel_{ij}$ in equations (\ref{consyn1c}) and
(\ref{hpara}), we can read off $\beta$:
\begin{equation}
\label{beta}
    \beta(\vec{x},\tau) = \int d^3k\,e^{i\vec{k}\cdot\vec{x}}
        \,{1\over 2k^2} \left\{ h(\vec{k},\tau) + 6\eta(\vec{k},\tau)
	\right\} \,.
\end{equation}
Then from equations (\ref{consyn2}), the conformal Newtonian potentials
$\phi$ and $\psi$ are related to the synchronous potentials $h$ and
$\eta$ in $k$-space by
\begin{eqnarray}
\label{trans2}
     \psi(\vec{k},\tau) &=& {1\over 2k^2} \left\{\ddot{h}(\vec{k},\tau)
	+ 6\ddot{\eta}(\vec{k},\tau) + {\dot{a} \over a} \left[
	\dot{h}(\vec{k},\tau) + 6\dot{\eta}(\vec{k},\tau)
        \right] \right\} \,,\nonumber\\
     \phi(\vec{k},\tau) &=&
         \eta(\vec{k},\tau) - {1\over 2k^2}{\dot{a} \over a}
         \left[ \dot{h}(\vec{k},\tau) + 6\dot{\eta}(\vec{k},\tau) \right] \,.
\end{eqnarray}
The other components of the metric perturbations, $w_i,
\sigma^\perp_{ij}\,$, and $\sigma^T_{ij}\,$, are zero in both gauges.

\section{Einstein Equations and Energy-Momentum Conservation}
For a homogeneous Friedmann-Robertson-Walker universe with energy
density $\bar{\rho}(\tau)$ and pressure $\bar{P}(\tau)$, the Einstein
equations give the following evolution equations for the expansion
factor $a(\tau)$:
\begin{eqnarray}
\label{friedmann}
	\left( {\dot{a}\over a} \right)^2 &=& {8\pi\over
		3}Ga^2 \bar{\rho} - \kappa \,,\\
	{d\over d\tau} \left( {\dot{a}\over a} \right)
		&=& -{4\pi\over 3}Ga^2 (\bar{\rho}+3\bar{P}) \,,
\label{friedmann2}
\end{eqnarray}
where the dots denote derivatives with respect to $\tau$, and $\kappa$
is positive, zero, or negative for a closed, flat, or open universe,
respectively.  We consider only $\Omega=1$ models in this paper, so we
set $\kappa=0$.  It follows that the expansion factor scales as
$a\propto\tau$ in the radiation-dominated era and $a\propto\tau^2$ in
the matter-dominated era.

We find it most convenient to solve the linearized Einstein equations
in the two gauges in the Fourier space $k$.  In the synchronous gauge,
the scalar perturbations are characterized by $h(\vec{k},\tau)$ and
$\eta(\vec{k},\tau)$ in equation (\ref{hijk}).  In terms of $h$ and
$\eta$, the time-time, longitudinal time-space, trace space-space, and
longitudinal traceless space-space parts of the Einstein equations
give the following four equations to linear order in $k$-space:
\medskip
\newline{\it Synchronous gauge ---\hfil}
\begin{mathletters}
\begin{eqnarray}
    k^2\eta - {1\over 2}{\dot{a}\over a} \dot{h}
        &=& 4\pi G a^2 \delta T^0{}_{\!0}(\mbox{Syn})
	\,,\label{ein-syna}\\
    k^2 \dot{\eta} &=& 4\pi Ga^2 (\bar{\rho}+\bar{P})
	\theta(\mbox{Syn})	 \,,\label{ein-synb}\\
    \ddot{h} + 2{\dot{a}\over a} \dot{h} - 2k^2
	\eta &=& -8\pi G a^2 \delta T^i{}_{\!i}(\mbox{Syn})
	\,,\label{ein-sync}\\
    \ddot{h}+6\ddot{\eta} + 2{\dot{a}\over a}\left(\dot{h}+6\dot{\eta}
	\right) - 2k^2\eta &=& -24\pi G a^2 (\bar{\rho}+\bar{P})
	\Theta(\mbox{Syn})	\,.\label{ein-synd}
\end{eqnarray}
\label{ein-syn}
\end{mathletters}
The label ``Syn'' is used to distinguish the components of the
energy-momentum tensor in the synchronous gauge from those in the
conformal Newtonian gauge.  The variables $\theta$ and $\Theta$ are
defined as
\begin{equation}
\label{theta}
 	(\bar{\rho}+\bar{P})\theta \equiv i k^j \delta T^0{}_{\!j}\,,
	\qquad	(\bar{\rho}+\bar{P})\Theta \equiv -(\hat{k}_i\hat{k}_j
	- {1\over 3} \delta_{ij})\Sigma^i{}_{\!j}\,,
\end{equation}
and $\Sigma^i{}_{\!j} \equiv T^i{}_{\!j}-\delta^i{}_{\!j} T^k{}_{\!k}/3$
denotes the traceless component of $T^i{}_{\!j}$.
When the different components of matter and radiation (i.e., CDM,
HDM, baryons, photons, and massless neutrinos) are treated
separately, $(\bar{\rho}+\bar{P})\theta = \sum_i
(\bar{\rho}_i+\bar{P}_i)\theta_i$ and $(\bar{\rho}+\bar{P})\Theta =
\sum_i (\bar{\rho}_i+\bar{P}_i)\Theta_i\,$, where the index $i$ runs
over the particle species.

In the conformal Newtonian gauge, the first-order perturbed Einstein
equations give
\medskip
\newline{\it Conformal Newtonian gauge ---\hfil}
\begin{mathletters}
\begin{eqnarray}
    k^2\phi + 3{\dot{a}\over a} \left( \dot{\phi} + {\dot{a}\over a}\psi
	\right) &=& 4\pi G a^2 \delta T^0{}_{\!0}(\mbox{Con}) \,,
	\label{ein-cona}\\
    k^2 \left( \dot{\phi} + {\dot{a}\over a}\psi \right)
	 &=& 4\pi G a^2 (\bar{\rho}+\bar{P}) \theta(\mbox{Con})
	 \,,\label{ein-conb}\\
    \ddot{\phi} + {\dot{a}\over a} (\dot{\psi}+2\dot{\phi})
	+\left(2{\ddot{a} \over a} - {\dot{a}^2 \over a^2}\right)\psi
	+ {k^2 \over 3} (\phi-\psi)
	&=& {4\pi\over 3} G a^2 \delta T^i{}_{\!i}(\mbox{Con})
	\,,\label{ein-conc}\\
    k^2(\phi-\psi) &=& 12\pi G a^2 (\bar{\rho}+\bar{P})\Theta(\mbox{Con})
	\,,\label{ein-cond}
\end{eqnarray}
\label{ein-con}
\end{mathletters}
where ``Con'' labels the conformal Newtonian coordinates.  Next we will
derive the transformation relating $\delta T^\mu{}_{\!\nu}$ in the two
gauges.

For a perfect fluid of energy density $\rho$ and pressure $P$,
the energy-momentum tensor has the form
\begin{equation}
	T^\mu{}_{\!\nu} = P g^\mu{}_{\!\nu} + (\rho+P) U^\mu U_\nu \,,
\end{equation}
where $U^\mu = dx^\mu /\sqrt{-ds^2} $ is the four-velocity of the
fluid.  The pressure $P$ and energy density $\rho$ of a perfect fluid
at a given point are defined to be the pressure and energy density
measured by a comoving observer at rest with the fluid at the instant
of measurements.  For a fluid moving with a small coordinate velocity
$v^i \equiv dx^i/d\tau$, $v^i$ can be treated as a perturbation of the
same order as $\delta\rho=\rho-\bar{\rho}$, $\delta P=P-\bar{P}$, and
the metric perturbations.  Then to linear order in the perturbations the
energy-momentum tensor is given by
\begin{eqnarray}
	T^0{}_{\!0} &=& -(\bar{\rho} + \delta\rho) \,,\nonumber\\
	T^0{}_{\!i} &=& (\bar{\rho}+\bar{P}) v_i  = -T^i{}_{\!0}\,,\nonumber\\
	T^i{}_{\!j} &=& (\bar{P} + \delta P) \delta^i{}_{\!j}
		+ \Sigma^i{}_{\!j} \,,\qquad
		\Sigma^i{}_{\!i}=0 \,,
\end{eqnarray}
where we have allowed an anisotropic shear perturbation
$\Sigma^i{}_{\!j}$ in $T^i{}_{\!j}$.  As we shall see, since the
photons are tightly coupled to the baryons before recombination, the
dominant contribution to this shear stress comes from the neutrinos.
We note that for a fluid, $\theta$ defined in equation (\ref{theta}) is
simply the divergence of the fluid velocity: $\theta = ik^j v_j$.

The energy-momentum tensor $T^\mu{}_{\!\nu}(\mbox{Syn})$ in
the synchronous gauge is related to $T^\mu{}_{\!\nu}(\mbox{Con})$ in
the conformal Newtonian gauge by the transformation
\begin{equation}
        T^\mu{}_{\!\nu}(\mbox{Syn}) =
	{\partial \hat{x}^\mu \over \partial x^\sigma}
        {\partial x^\rho \over \partial \hat{x}^\nu}
	T^\sigma{}_{\!\rho}(\mbox{Con})\,,
\end{equation}
where $\hat{x}^\mu$ and $x^\mu$ denote the synchronous and the
conformal Newtonian coordinates, respectively.  It follows that to
linear order, $T^0{}_{\!0}(\mbox{Syn})= T^0{}_{\!0}(\mbox{Con})\,,
T^0{}_{\!j}(\mbox{Syn})= T^0{}_{\!j}(\mbox{Con})+ik_j\alpha
(\bar{\rho}+\bar{P})\,$, and
$T^i{}_{\!j}(\mbox{Syn})=T^i{}_{\!j}(\mbox{Con})\,$, where $\alpha =
\hat{x}^0 - x^0 = (\dot{h}+6\dot{\eta})/2k^2$ in $k$-space from
equations (\ref{alpha}) and (\ref{beta}).  Let $\delta \equiv
\delta\rho/\bar{\rho}=-\delta T^0{}_{\!0}/\bar{\rho}$.  Evaluating
the perturbations at the same spacetime coordinate values, we obtain
\begin{mathletters}
\begin{eqnarray}
	\delta(\mbox{Syn}) &=& \delta(\mbox{Con})
		- \alpha {\dot{\bar\rho} \over {\bar\rho}}\,,\\
	\theta(\mbox{Syn}) &=& \theta(\mbox{Con}) - \alpha k^2\,,\\
	\delta P(\mbox{Syn}) &=& \delta P(\mbox{Con})
		-\alpha\dot{\bar{P}} \,,\\
	\Theta(\mbox{Syn}) &=& \Theta(\mbox{Con}) \,.
\end{eqnarray}
\label{deltat}
\end{mathletters}
This transformation also applies to individual species when more than one
particle species contributes to the energy-momentum tensor, provided that
the appropriate $\bar{\rho}$ and $\bar{P}$ are used for each component.

The non-relativistic fluid description is appropriate for the CDM and
the baryon components.  The photon and the neutrino components,
however, can be appropriately described only by their full
distribution functions in phase space.  The energy-momentum tensor in
this case is expressed through integrals over momenta of the
distribution functions.  We will discuss it in detail in \S 5.

The conservation of energy-momentum is a consequence of the Einstein
equations.  Let $w \equiv P/\rho$ describe the equation of state.
Then the perturbed part of energy-momentum conservation equations
\begin{equation}
\label{Econs}
	T^{\mu\nu}{}_{\!;\mu} = \partial_\mu T^{\mu\nu}
	+ \Gamma^\nu{}_{\!\alpha\beta} T^{\alpha\beta}
	+ \Gamma^\alpha{}_{\!\alpha\beta} T^{\nu\beta} = 0
\end{equation}
in $k$-space implies
\medskip
\newline{\it Synchronous gauge ---\hfil}
\begin{eqnarray}
\label{fluid}
	\dot{\delta} &=& - (1+w) \left(\theta+{\dot{h}\over 2}\right)
	  - 3{\dot{a}\over a} \left({\delta P \over \delta\rho} - w
	  \right)\delta  \,,\nonumber\\
	\dot{\theta} &=& - {\dot{a}\over a} (1-3w)\theta - {\dot{w}\over
	     1+w}\theta + {\delta P/\delta\rho \over 1+w}\,k^2\delta
	     - k^2 \Theta\,,
\end{eqnarray}
\newline{\it Conformal Newtonian gauge ---\hfil}
\begin{eqnarray}
\label{fluid2}
	\dot{\delta} &=& - (1+w) \left(\theta-3{\dot{\phi}}\right)
	  - 3{\dot{a}\over a} \left({\delta P \over \delta\rho} - w
	  \right)\delta \,,\nonumber\\
	\dot{\theta} &=& - {\dot{a}\over a} (1-3w)\theta - {\dot{w}\over
	     1+w}\theta + {\delta P/\delta\rho \over 1+w}\,k^2\delta
	     - k^2 \Theta + k^2 \psi \,.
\end{eqnarray}
These equations are valid for a single uncoupled fluid, or for the net
(mass-averaged) $\delta$ and $\theta$ for all fluids.  They need to be
modified for individual components if the components interact with each
other.  An example is the baryonic fluid in our model, which couples
to the photons before recombination via Thomson scattering.
In the next section we will show that an extra term representing
momentum transfer between the two components needs to be added to the
$\dot\delta$ equation for the baryons.

For the isentropic primordial perturbations considered in this paper,
the equations above simplify since $\delta P/\delta\rho = c^2_s = w$,
where $c_s$ is the adiabatic sound speed in the fluid.  Although
entropy is generated from the coupling of baryons and photons before
recombination, it is a first-order perturbation and only appears in
the energy-momentum conservation equations in the second order.  Thus
the terms proportional to $(\delta P/\delta\rho - w)$ above can be
neglected when the equations are applied to CDM and baryons in the
next section.  Even in the case of isocurvature models, which may have
large entropy perturbations ab initio, these terms are generally
small.

\section{Evolution Equations for Matter and Radiation}
\label{sec:boltz}
\subsection{Phase Space and the Boltzmann Equation}
A phase space is described by six variables: three positions $x^i$ and
their conjugate momenta $P_i$.  Our treatment of phase space is based
on the time-slicing of a definite gauge (synchronous or conformal
Newtonian).  Although this approach is not manifestly covariant, it
yields correct results provided we convert gauge-dependent quantities
to observables at the end of our computation.

The conjugate momentum has the property that it is simply the spatial
part of the 4-momentum with lower indices, i.e., for a particle of mass
$m$, $P_i=mU_i\,,$ where $U_i=dx_i/\sqrt{-ds^2}$.  One can verify that
the conjugate momentum is related to the proper momentum $p^i=p_i$ measured
by an observer at a fixed spatial coordinate value by
\begin{eqnarray}
 	P_i &=a(\delta_{ij}+\frac{1}{2}h_{ij}) p^j\,,\qquad
	 &\mbox{in synchronous gauge}\,, \nonumber\\
	P_i &=a(1-\phi)p_i\,,\qquad
	 &\mbox{in conformal Newtonian gauge}\,.
\end{eqnarray}
In the absence of metric perturbations, Hamilton's
equations imply that the conjugate momenta are constant, so the proper
momenta redshift as $a^{-1}$.

The phase space distribution of the particles gives the
number of particles in a differential volume $dx^1 dx^2 dx^3 dP_1 dP_2
dP_3$ in phase space:
\begin{equation}
	f(x^i,P_j,\tau)dx^1 dx^2 dx^3 dP_1 dP_2 dP_3 = dN \ .
\end{equation}
Importantly, $f$ is a scalar and is invariant under canonical
transformations.  The zeroth-order phase space distribution is the
Fermi-Dirac distribution for fermions ($+$ sign) and the Bose-Einstein
distribution for bosons ($-$ sign):
\begin{equation}
   f_0=f_0(\epsilon) = {g_s\over h_{\rm P}^3}{1\over e^{\epsilon/
     k_{\rm B} T_0} \pm 1}\,,
\end{equation}
where $\epsilon = a(p^2+m^2)^{1/2} =  (P^2 + a^2m^2)^{1/2}\,$, $T_0 = a T$
denotes the temperature of the particles today, the factor $g_s$ is the
number of spin degrees of freedom, and $h_{\rm P}$ and $k_{\rm B}$ are
the Planck and the Boltzmann constants.

When the spacetime is perturbed, $x^i$ and $P_i$ remain canonically
conjugate variables, with equations of motion given by Hamilton's
equations (Bertschinger 1993).  However, following common practice
(e.g., Bond \& Szalay 1983) we shall find it convenient to replace
$P_j$ by $q_j\equiv ap_j$ in order to eliminate the metric perturbations
from the definition of the momenta.  Moreover, we shall write the
comoving 3-momentum $q_j$ in terms of its magnitude and direction:
$q_j=qn_j$ where $n^in_i=\delta_{ij}n^in^j=1$.  Thus, we change our
phase space variables, replacing $f(x^i,P_j,\tau)$ by $f(x^i,q,n_j,\tau)$.
While this is not a canonical transformation (i.e., $q_i$ is not the
momentum conjugate to $x^i$), it is perfectly valid provided that we
correctly transform the momenta in Hamilton's equations.  Note that we
do not transform $f$.  Because $q_j$ are not the conjugate momenta,
$d^3xd^3q$ is not the phase space volume element, and $fd^3xd^3q$ is
not the particle number.  In the conformal Newtonian gauge, for example,
$(1-3\phi)fd^3xd^3q$ is the particle number; this result is sensible
because $a(1-\phi)dx^i$ is the proper distance.

In the perturbed case we shall continue to define $\epsilon$ as $a(\tau)$
times the proper energy measured by a comoving observer,  $\epsilon=
(q^2+a^2m^2)^{1/2}$.  This is related to the time component of the
4-momentum by $P_0=-\epsilon$ in the synchronous gauge and $P_0=-(1+\psi)
\epsilon$ in the conformal Newtonian gauge.  For the CDM+HDM model we
are interested in, the photons, the massless neutrinos, and the 7 eV
neutrinos at the time of neutrino decoupling are all ultra-relativistic
particles, so $\epsilon$ in the unperturbed Fermi-Dirac and Bose-Einstein
distributions can be simply replaced by the new variable $q$.

The general expression for the energy-momentum tensor written
in terms of the distribution function and the 4-momentum components
is given by
\begin{equation}
\label{tmunu}
	T_{\mu\nu}= \int dP_1 dP_2 dP_3\,(-g)^{-1/2}\,
	{P_\mu P_\nu\over P^0} f(x^i,P_j,\tau) \,,
\end{equation}
where $g$ denotes the determinant of $g_{\mu\nu}$.
It is convenient to write the phase space distribution as a
zeroth-order distribution plus a perturbed piece in
the new variables $q$ and $n_j$:
\begin{equation}
	f(x^i,P_j,\tau) = f_0(q) \left[ 1 + \Psi(x^i,q,n_j,\tau) \right] \,.
\end{equation}

In the synchronous gauge, we have $(-g)^{-1/2} = a^{-4}(1-\frac{1}{2}h)$
and $dP_1 dP_2 dP_3 = (1+\frac{3}{2}h_{ij}n_in_j) q^2 dq d\Omega$
to linear order, where $h \equiv h_{ii}$ and $d\Omega$ is the solid
angle associated with direction $n_i$.
Using the relations $\int d\Omega\,n_i n_j = 4\pi \delta_{ij}/3$ and
$\int d\Omega\,n_i = \int d\Omega\,n_i n_j n_k = 0$, it then follows
from equation (\ref{tmunu}) that
\begin{eqnarray}
\label{tmunu2}
	T^0{}_{\!0} &=& - a^{-4} \int q^2dq\,d\Omega\,
		\sqrt{q^2+m^2a^2}\,f_0(q)\,(1+\Psi) \,,\nonumber\\
	T^0{}_{\!i} &=& a^{-4} \int q^2dq\,d\Omega\,
		q\,n_i\,f_0(q)\,\Psi \,,\\
	T^i{}_{\!j} &=& a^{-4} \int q^2dqd\Omega
		\,{q^2 n_i n_j\over \sqrt{q^2+m^2a^2}}\,f_0(q)\,(1+\Psi)
		\nonumber
\end{eqnarray}
to linear order in the perturbations.  Note that we have eliminated the
explicit dependence on the metric perturbations in equation (\ref{tmunu})
by redefining $P_i$ in terms of $q$ and $n_i$.   Note also that the
comoving energy $\epsilon(q,\tau)=(q^2+a^2m^2)^{1/2}$ is used in the
integrands but not in the argument of the unperturbed distribution function.

In the conformal Newtonian gauge, $(-g)^{-1/2} = a^{-4}(1-\psi+3\phi)$
and $dP_1 dP_2 dP_3 = (1-3\phi) q^2 dq d\Omega$.  It then follows that
the components of the energy-momentum tensor have the
same form as in equations (\ref{tmunu2}).  Of course, it is understood that
the variables $q$ and $n_i$ in this case are defined in relation to
the conjugate momentum $P_i$ in the conformal Newtonian coordinates,
and not the synchronous coordinates.  (They differ because comoving
observers in the two coordinate systems are not the same.) The expansion
factor $a$ and $\Psi$ are evaluated at the coordinates $(x^i,\tau)$
in the conformal Newtonian gauge.

The phase space distribution evolves according to the Boltzmann equation.
In terms of our variables $(x^i,q,n_j,\tau)$ this is
\begin{equation}
	{Df \over d\tau} = {\partial f \over \partial \tau}
	+ {dx^i \over d\tau}{\partial f\over \partial x^i}
	+ {dq \over d\tau}{\partial f\over \partial q}
	+ {dn_i \over d\tau}{\partial f\over \partial n_i}
	= \left( {\partial f \over \partial\tau} \right)_C\,,
\end{equation}
where the right-hand side involves terms due to collisions,
whose form depends on the type of particle interactions involved.
 From the geodesic equation
\begin{equation}
	P^0 {dP^\mu \over d\tau} + \Gamma^\mu{}_{\!\alpha\beta}
	\,P^\alpha P^\beta = 0 \,,
\end{equation}
it is straightforward to show that
\begin{eqnarray}
	dq/d\tau &=& -{1\over 2} q\dot{h}_{ij}n_i n_j \qquad \mbox{in
	synchronous gauge}\,, \nonumber\\
	dq/d\tau &=& q\dot{\phi}-\epsilon(q,\tau)\,n_i \partial_i \psi \qquad
	\mbox{in conformal Newtonian gauge}\,,
\end{eqnarray}
and $dn_i/d\tau$ is $O(h)$.  Since $\partial f/\partial n_i$
is also a first-order quantity, the term $(d n_i/d\tau)(\partial
f/\partial n_i)$ in the Boltzmann equation can be neglected to first
order.  Then the Boltzmann equation in $k$-space can be written as
\medskip
\newline {\it Synchronous gauge ---\hfil}
\begin{equation}
\label{bolt-syn}
  {\partial \Psi \over \partial \tau} + i\,{q\over\epsilon}
	(\vec{k}\cdot \hat{n})\Psi +
     {d\ln f_0 \over d\ln q}\, \left[\dot{\eta} - {\dot{h}+6\dot{\eta}
       \over 2}(\hat{k}\cdot\hat{n})^2 \right] =
     {1\over f_0}\,\left( {\partial f \over \partial\tau} \right)_C \,,
\end{equation}
\newline {\it Conformal Newtonian gauge ---\hfil}
\begin{equation}
\label{bolt-con}
  {\partial \Psi \over \partial \tau} + i\,{q \over
	\epsilon}\,(\vec{k}\cdot \hat{n})\,\Psi +
     {d\ln f_0 \over d\ln q} \left[\dot{\phi} - i\,{\epsilon\over q}
       (\vec{k}\cdot\hat{n})\,\psi \right] =
   {1\over f_0}\,\left( {\partial f \over \partial\tau} \right)_C \,.
\end{equation}
Equations (\ref{bolt-syn}) and (\ref{bolt-con}) can also be derived using
the canonical phase space variables $x^i$ and $P_j$ and Hamilton's equations
(instead of the geodesic equation), followed by a transformation from $P_j$
to $qn_j$.

The terms in the Boltzmann equation depend on the direction of the
momentum $\hat{n}$ only through its angle with $\vec{k}$.  (We shall see
that this is true of the collision term for photons as well as
the convective and metric perturbation terms.) Therefore, if the
momentum-dependence of the initial phase space perturbation is axially
symmetric about $\vec k$, it will remain axially symmetric.
If axially-asymmetric perturbations in the neutrinos or other
collisionless particles are produced, they would generate no scalar
metric perturbations and thus would have no effect on other species.
Therefore, we shall assume that the initial momentum-dependence is
axially symmetric so that $\Psi$ depends on $\vec{q}=q\hat{n}$ only
through $q$ and $\hat{k}\cdot\hat{n}$.  This assumption, which
effectively reduces the dimensionality of phase space perturbations by
one (after Fourier transforming on the spatial coordinates), has been
made (implicitly, if not explicitly) in all previous studies of the
evolution of scalar perturbations.

\subsection{Cold Dark Matter}
CDM interacts with other particles only through gravity and can be
treated as a pressureless perfect fluid.  The CDM particles can be
used to define the synchronous coordinates and therefore have zero
peculiar velocities in this gauge.  Setting $\theta=\Theta=0$ and
$w=\dot{w}=c_s^2=0$ in equations (\ref{fluid}) leads to
\medskip
\newline {\it Synchronous gauge ---\hfil}
\begin{equation}
\label{cdm}
	\dot{\delta_c} = -\frac{1}{2}\,\dot{h} \,.
\end{equation}
The CDM fluid velocity in the conformal Newtonian gauge, however, is
not zero in general.  In $k$-space, equations (\ref{fluid2}) give
\medskip
\newline {\it Conformal Newtonian gauge ---\hfil}
\begin{equation}
\label{cdm2}
	\dot{\delta_c} = -\theta_c + 3\dot{\phi}\,, \quad
	\dot{\theta}_c = - {\dot{a}\over a}\,\theta_c+k^2\psi \,.
\end{equation}
The subscript $c$ in $\delta_c$ and $\theta_c$ denotes the cold dark matter.

\subsection{Massless Neutrinos}
\label{sec:lessnu}
The energy density and the pressure for massless neutrinos
(labeled by subscripts $\nu$) are $\rho_\nu = 3P_\nu = -T^0{}_{\!0} =
T^i{}_{\!i}$.  From equations (\ref{tmunu2}) the unperturbed
energy density $\bar{\rho}_\nu$ and pressure $\bar{P}_\nu$ are given by
\begin{equation}
	 \bar{\rho}_\nu = 3 \bar{P}_\nu = a^{-4}
	\int q^2dq d\Omega\,q f_0(q) \,,
\end{equation}
and the perturbations of energy density $\delta\rho_\nu$, pressure $\delta
P_\nu$, energy flux $\delta T^0_{\nu\,i}\,$, and shear stress
$\Sigma^i_{\nu\,j}=T^i_{\nu\,j}-P_\nu\delta_{ij}$ are given by
\begin{eqnarray}
     \delta\rho_\nu &=& 3 \delta P_\nu = a^{-4}
        \int q^2dq d\Omega\,q f_0(q) \Psi \,,\nonumber\\
     \delta T^0_{\nu\,i} &=& a^{-4}
	\int q^2dq d\Omega\,qn_i\,f_0(q) \Psi \,,\\
     \Sigma^i_{\nu\,j} &=& a^{-4}
	\int q^2dq d\Omega\,q(n_in_j-\frac{1}{3}\delta_{ij})\,f_0(q)\Psi \,.
	\nonumber
\end{eqnarray}
The unperturbed energy flux and shear stress are zero.

The Boltzmann equation simplifies for massless particles, for which
$\epsilon=q$.  To reduce the number of variables we integrate out the
$q$-dependence in the neutrino distribution function and expand the
angular dependence of the perturbation in a series of Legendre polynomials
$P_l(\hat{k}\cdot\hat{n})$:
\begin{equation}
\label{fsubl}
      F_\nu(\vec{k},\hat{n},\tau) \equiv {\int q^2 dq\,q f_0(q)\Psi
	\over \int q^2 dq\,q f_0(q)} \equiv \sum_{l=0}^\infty(-i)^l
	F_{\nu\,l}(\vec{k},\tau)P_l(\hat{k}\cdot\hat{n})\,.
\end{equation}
As noted in \S 5.1, the dependence on $\vec n$ arises only through
$\hat{k}\cdot\hat{n}$, so that a general distribution may be represented
as in equation (\ref{fsubl}).

In terms of the new variable $F_\nu(\vec k,\vec n,\tau)$ and its harmonic
expansion coefficients, the perturbations $\delta_\nu$,
$\bar{\rho}_\nu$, $\theta_\nu$, and $\Theta_\nu$  (defined in eq.
\ref{theta}) take the form
\begin{eqnarray}
      \delta_\nu &=& {1\over 4\pi} \int d\Omega
		F_\nu(\vec{k},\hat{n},\tau)=F_{\nu\,0}\,,\nonumber\\
      \theta_\nu &=& {3i \over 16\pi} \int d\Omega\,
		(\vec{k}\cdot\hat{n}) F_\nu(\vec{k},\hat{n},\tau)
		={1\over 4} k F_{\nu\,1}\,,\\
      \Theta_\nu &=& -{3 \over 16\pi} \int d\Omega
	\left[(\hat{k}\cdot\hat{n})^2 - {1\over 3}\right]
	 F_\nu(\vec{k},\hat{n},\tau)={1\over 10} F_{\nu\,2}\,,\nonumber
\end{eqnarray}
where we have used $\rho_\nu=3P_\nu$ for the massless neutrinos.

Integrating equations (\ref{bolt-syn}) and (\ref{bolt-con}) over
$q^2dq\,q f_0(q)$ and dividing them by $\int q^2dq\,q f_0(q)$, the
Boltzmann equation for massless neutrinos becomes
\begin{eqnarray}
\label{bolmn}
  {\partial F_\nu\over\partial\tau}+ik\mu F_\nu &=&
    -\frac{2}{3}\dot h-\frac{4}{3}(\dot h+6\dot\eta)P_2(\mu)\qquad
      \mbox{in synchronous gauge}\,, \nonumber\\
  {\partial F_\nu\over\partial\tau}+ik\mu F_\nu &=&
    4\,(\dot\phi-ik\mu\psi)\qquad
      \mbox{in conformal Newtonian gauge}\,,
\end{eqnarray}
where $\mu\equiv\hat k\cdot\hat n$ and $P_2(\mu)=\frac{1}{2}(3\mu^2-1)$
is the Legendre polynomial of degree 2.  Substituting the Legendre
expansion for $F_\nu$ and using the orthonormality of the Legendre
polynomials and the recursion relation $(l+1)P_{l+1}(\mu) = (2l+1)
\mu P_l(\mu) -l P_{l-1}(\mu)$, we obtain
\medskip
\newline {\it Synchronous gauge ---\hfil}
\begin{eqnarray}
\label{massless}
	\dot{\delta}_\nu &=& -{4\over 3}\theta_\nu
		-{2\over 3}\dot{h} \,,\nonumber\\
	\dot{\theta}_\nu &=& k^2 \left(\frac{1}{4}\delta_\nu
		- \Theta_\nu \right)\,,\nonumber\\
	\dot{F}_{\nu\,2} &=& 10\dot\Theta_\nu = {8\over 3}\theta_\nu
	    	- {3\over 7} k F_{\nu\,3} + {4\over 3}\dot{h}
	  	+ 8 \dot{\eta} \,,\nonumber\\
	\dot{F}_{\nu\,l} &=& k\left[ {l\over 2l-1}
	 	F_{\nu\,(l-1)} - {l+1 \over 2l+3}
		F_{\nu\,(l+1)} \right]\,, \quad l \geq 3 \,.
\end{eqnarray}
\newline {\it Conformal Newtonian gauge ---\hfil}
\begin{eqnarray}
\label{massless2}
	\dot{\delta}_\nu &=& -{4\over 3}\theta_\nu
		+4\dot{\phi} \,, \nonumber\\
	\dot{\theta}_\nu &=&k^2 \left(\frac{1}{4}\delta_\nu
		- \Theta_\nu \right)
		+ k^2\psi \,, \nonumber\\
	\dot{F}_{\nu\,l} &=& k\left[ {l\over 2l-1}
 		F_{\nu\,(l-1)} - {l+1 \over 2l+3}
		F_{\nu\,(l+1)} \right]\,, \quad l \geq 2 \,.
\end{eqnarray}
This set of equations governs the evolution of the phase space
distribution of massless neutrinos.  Note that a given mode $F_l$ is
coupled only to the $(l-1)$ and $(l+1)$ neighboring modes.

\subsection{Massive Neutrinos}

Massive neutrinos also obey the collisionless Boltzmann equation.
The evolution of the distribution function for massive neutrinos
is, however, complicated by their nonzero mass.  From equations
(\ref{tmunu2}), the unperturbed energy density and pressure for
massive neutrinos (labeled by subscripts ``$h$'' for HDM) are given by
\begin{equation}
	\bar{\rho}_h = a^{-4} \int q^2dq\,d\Omega\,
		\epsilon f_0(q) \,,\qquad
	\bar{P}_h = {1\over 3} a^{-4} \int q^2dq\,d\Omega\,
		{q^2\over\epsilon} f_0(q) \,,
\end{equation}
where $\epsilon = \epsilon(q,\tau) = \sqrt{q^2 + m_\nu^2a^2}$, while
the perturbations are
\begin{eqnarray}
\label{delta}
   \delta\rho_h =& a^{-4} \int q^2dq\,d\Omega\,\epsilon f_0(q) \Psi
	\,,\qquad  \delta P_h =& {1\over 3} a^{-4}
        \int q^2dq\,d\Omega\,{q^2\over \epsilon} f_0(q) \Psi \,,\nonumber\\
     \delta T^0_{h\,i} =& a^{-4}
	\int q^2dq\,d\Omega\,q n_i\,f_0(q) \Psi \,,\qquad
     \Sigma^i_{h\,j} =& a^{-4}
	\int q^2dq\,d\Omega\,{q^2\over\epsilon}
	(n_in_j-{1\over 3}\delta_{ij})\,f_0(q) \Psi \,.
\end{eqnarray}
Since the comoving energy-momentum relation $\epsilon(q,\tau)$ depends
on both the momentum and time, we can not simplify the calculations by
integrating out the $q$-dependence in the distribution function as we
did for the massless neutrinos above (see eq. \ref{fsubl}).  Instead of
applying equation (\ref{fsubl}), we expand the perturbation $\Psi$
directly in a Legendre series
\begin{equation}
      \Psi(\vec{k},\hat{n},q,\tau)
	= \sum_{l=0}^\infty (-i)^l \Psi_l(\vec{k},q,\tau)
	P_l(\hat{k}\cdot\hat{n})\,.
\end{equation}
Then the perturbed energy density, pressure, energy flux, and shear
stress in $k$-space are given by
\begin{eqnarray}
\label{delta2}
   \delta\rho_h &=& 4\pi a^{-4}
        \int q^2 dq\,\epsilon f_0(q) \Psi_0 \,, \nonumber\\
   \delta P_h &=& {4\pi \over 3} a^{-4}
        \int q^2 dq\,{q^2\over \epsilon} f_0(q) \Psi_0
	 \,, \nonumber\\
   (\bar\rho_h +\bar P_h) \theta_h &=& {4\pi \over 3} k a^{-4}
	\int q^2 dq\,q f_0(q)\Psi_1 \,, \nonumber\\
   (\bar\rho_h +\bar P_h) \Theta_h &=& {8\pi\over 15} a^{-4}
	\int q^2 dq\,{q^2\over\epsilon} f_0(q) \Psi_2 \,.
\end{eqnarray}

Following the same procedure used for the massless neutrinos, the Boltzmann
equation becomes
\medskip
\newline{\it Synchronous gauge ---\hfil}
\begin{eqnarray}
\label{massive}
     \dot{\Psi}_0 &=& -{qk\over 3\epsilon}\Psi_1
  	  +{1\over 6}\dot{h} {d\ln f_0\over d\ln q}
		\,, \nonumber\\
     \dot{\Psi}_1 &=& {qk\over\epsilon} \left(\Psi_0
		      - {2\over 5} \Psi_2 \right) \,, \nonumber\\
     \dot{\Psi}_2 &=& {qk\over\epsilon} \left(
	{2\over 3}\Psi_1 - {3\over 7} \Psi_3 \right)
	 - \left( {1\over 3}\dot{h} + 2 \dot{\eta} \right)
	{d\ln f_0\over d\ln q} \,,\\
    \dot{\Psi}_l &=& {qk \over \epsilon} \left( {l\over 2l-1}\Psi_{l-1}
        - {l+1 \over 2l+3} \Psi_{l+1} \right)\,,
	\quad l \geq 3 \,. \nonumber
\end{eqnarray}
\newline {\it Conformal Newtonian gauge ---\hfil}
\begin{eqnarray}
\label{massive2}
     \dot{\Psi}_0 &=& -{qk\over 3\epsilon}\Psi_1
  	  -\dot{\phi} {d\ln f_0\over d\ln q} \,, \nonumber\\
     \dot{\Psi}_1 &=& {qk\over\epsilon} \left(\Psi_0
	  - {2\over 5} \Psi_2 \right)
	  - {\epsilon\,k\over q} \psi {d\ln f_0\over d\ln q} \,,\\
    \dot{\Psi}_l &=& {qk \over \epsilon} \left(
	{l\over 2l-1} \Psi_{l-1} - {l+1 \over 2l+3} \Psi_{l+1} \right)\,,
	\quad l \geq 2 \,. \nonumber
\end{eqnarray}
Because of the $q$-dependence in these equations, it requires much
more computing time to carry out the time integration for the massive
neutrino.  Bond \& Szalay (1983) used a 16-point Gauss-Legendre method
to approximate the $q$-integration.  We do not use this method and
instead perform the integration using cubic splines (plus a remainder
obtained by asymptotic expansion) with a $q$-grid of 100 $q$-points
for every wavenumber $k$.  We verified that this was enough to ensure
a relative accuracy no worse than $10^{-6}$ by trying the integration
with 200 points.  Then the perturbations $\delta\rho_h$, $\delta P_h$,
$\theta_h$, and $\Theta_h$ that enter the right-hand side of the
Einstein equations are calculated from equations (\ref{delta2}) by
numerically integrating $\Psi_0, \Psi_1$ and $\Psi_2$ over $q$.

\subsection{Photons}
Photons evolve differently before and after recombination.  Before
recombination, photons and baryons are tightly coupled, interacting
mainly via Thomson scattering (and the electrostatic coupling of electrons
and ions).  In Thomson scatterings, the photon energy $h\nu$ is assumed
to be much less than the electron rest mass $m_e \sim 0.511$ MeV and
the recoil of the electron in the initial electron rest frame is
neglected.  (We are concerned with the period after neutrino decoupling,
when $T < m_e$.) The classical differential cross section for Thomson
scattering is given by $d\sigma_T/d\Omega = 3\sigma_T(1+\cos^2\theta)
/16\pi$, where $\sigma_T = 0.6652 \times 10^{-24}$ cm$^2$ and $\theta$
is the scattering angle (e.g., Jackson 1975).  After recombination,
the universe gradually becomes transparent to radiation and photons
travel almost freely, although Thomson scattering continues to
transfer energy and momentum between the photons and the matter.

The evolution of the photon distribution function can be treated
in a similar way as the massless neutrinos, with the exception that
the collisional terms on the right-hand side of the Boltzmann equation
are now present.  We shall denote by $F_\gamma(\vec k,\hat{n},\tau)$
the momentum-averaged phase space density perturbation, defined as
in equation (\ref{fsubl}).  The linearized collision operator for Thomson
scattering is (Wilson \& Silk 1980; Dodelson \& Jubas 1993)
\begin{equation}
	\left( {\partial F_\gamma \over \partial\tau} \right)_C
	    = a n_e \sigma_T \left[ -F_\gamma(\vec{k},\hat{n},\tau)
	    + F_{\gamma\,0}(\vec{k},\tau) + 4\hat{n}\cdot\vec{v}_e
	    - {1\over 10} F_{\gamma\,2} P_2(\hat{k}\cdot\hat{n})
     \right] \,,
\end{equation}
where $n_e$ and $\vec{v}_e$ are the proper mean density and
velocity of the electrons.  The term proportional to $P_2$ comes from
the angular dependence $\cos^2\theta$ in the Thomson cross section
above.  The $P_2$ term was neglected by Peebles \& Yu (1970), but
was included by Wilson \& Silk (1980, 1981).  Expanding $F_\gamma
(\vec{k},\hat{n},\tau)$ in Legendre series as in equation (\ref{fsubl})
and using the relations $\hat{n}\cdot\vec{v}_e = i\theta_b P_1(\hat{k}
\cdot\hat{n})$, $F_{\gamma\,1}=k\theta_\gamma/4$, and $F_{\gamma\,2}=
10\Theta_\gamma$, the collision operator can be rewritten as
\begin{equation}
	\left( {\partial F_\gamma \over \partial\tau} \right)_C
	    = a n_e \sigma_T \left[ {4i\over k}
	    (\theta_\gamma-\theta_b)P_1 + 9\Theta_\gamma P_2
	    - \sum_{l\ge 3}^\infty (-i)^l F_{\gamma\,l} P_l \right]\,.
\end{equation}
The left-hand-side of the Boltzmann equation remains the same as for
the massless neutrinos, so we obtain
\medskip
\newline {\it Synchronous gauge ---\hfil}
\begin{eqnarray}
\label{photon}
     \dot{\delta}_\gamma &=& -{4\over 3}\theta_\gamma
	-{2\over 3}\dot{h} \,,\nonumber\\
     \dot{\theta}_\gamma &=& k^2 \left(\frac{1}{4}\delta_\gamma
	- \Theta_\gamma\right)
	+ a n_e \sigma_T (\theta_b - \theta_\gamma) \,,\nonumber\\
     \dot{F}_{\gamma\,2} &=& 10\dot\Theta_\gamma={8\over 3}\theta_\gamma
       - {3\over 7} k F_{\gamma\,3} + {4\over 3}\dot{h}
       + 8 \dot{\eta} - 9 an_e \sigma_T \Theta_\gamma
	\,, \nonumber\\
     \dot{F}_{\gamma\,l} &=& k\left[ {l\over 2l-1}
	F_{\gamma\,(l-1)} - {l+1 \over 2l+3} F_{\gamma\,(l+1)}\right]
	- an_e \sigma_T F_{\gamma\,l} \,,\quad l \geq 3 \,,
\end{eqnarray}
\newline {\it Conformal Newtonian gauge ---\hfil}
\begin{eqnarray}
\label{photon2}
     \dot{\delta}_\gamma &=& -{4\over 3}\theta_\gamma
	+4\dot{\phi} \,,\nonumber\\
     \dot{\theta}_\gamma &=& k^2 \left(\frac{1}{4}\delta_\gamma
	- \Theta_\gamma \right) + k^2 \psi
	+ a n_e \sigma_T (\theta_b-\theta_\gamma) \,,\nonumber\\
     \dot{F}_{\gamma\,2} &=& 10\dot\Theta_\gamma= {8\over 3}\theta_\gamma
        - {3\over 7} k F_{\gamma\,3}
	- 9 an_e \sigma_T \Theta_\gamma \,, \nonumber\\
     \dot{F}_{\gamma\,l} &=& k\left[ {l\over 2l-1}
	F_{\gamma\,(l-1)} - {l+1 \over 2l+3} F_{\gamma\,(l+1)}
	\right] - an_e \sigma_T F_{\gamma\,l} \,,\quad l \geq 3 \,.
\end{eqnarray}
The subscripts $\gamma$ and $b$ refer to photons and baryons
respectively.

\subsection{Baryons}
The baryons (and electrons) behave like a non-relativistic fluid described,
in the absence of coupling to radiation, by the energy-momentum conservation
equations (\ref{fluid}) and (\ref{fluid2}) with $\delta P/\delta\rho=
c_s^2=w\ll1$ and $\Theta=0$.  Since the baryons are very nonrelativistic
after neutrino decoupling (the period of interest), we may neglect $w$
and $\delta P/\delta\rho$ in all terms except the acoustic term $c_s^2k^2
\delta$ (which is important for sufficiently high $k$; note that the shear
stress term $k^2\Theta$ is far smaller so we neglect it).  Before
recombination, however, the coupling of the baryons and the photons
causes a transfer of momentum (and energy) between the two components.

 From equation (\ref{theta}) the momentum density $T^0{}_{\!j}$ for a given
species is related to $\theta\,$ by $ik^j \delta T^0{}_{\!j} =
(\bar{\rho}+\bar{P})\theta$.  The momentum transfer into the
photon component is represented by $an_e\sigma_T(\theta_b -
\theta_\gamma)$ of equations (\ref{photon}) and (\ref{photon2}).
Momentum conservation in Thomson scattering then implies that a term
$(4\bar\rho_\gamma/3
\bar\rho_b)\,an_e\sigma_T(\theta_\gamma - \theta_b)$ has to be added
to the equation for $\dot{\theta}_b\,$ (where we have used $\bar{P}_b
\ll \bar{\rho}_b$), so equations (\ref{fluid}) and (\ref{fluid2}) are
modified to become \medskip
\newline {\it Synchronous gauge ---\hfil}
\begin{eqnarray}
\label{baryon}
	\dot{\delta}_b &=& -\theta_b - {1\over 2}\dot{h} \,, \nonumber\\
	\dot{\theta}_b &=& -{\dot{a}\over a}\theta_b
	+ c_s^2 k^2\delta_b + {4\bar\rho_\gamma \over 3\bar\rho_b}
	 an_e\sigma_T (\theta_\gamma-\theta_b)\,,
\end{eqnarray}
\newline {\it Conformal Newtonian gauge ---\hfil}
\begin{eqnarray}
\label{baryon2}
	\dot{\delta}_b &=& -\theta_b + 3\dot{\phi} \,, \nonumber\\
	\dot{\theta}_b &=& -{\dot{a}\over a}\theta_b
	+ c_s^2 k^2\delta_b + {4\bar\rho_\gamma \over 3\bar\rho_b}
	 an_e\sigma_T (\theta_\gamma-\theta_b) + k^2\psi\,.
\end{eqnarray}

\subsection{Tight-Coupling Approximation}

Before recombination the Thomson opacity is so large that photons and
baryons are tightly coupled, with $an_e\sigma_T\equiv\tau_c^{-1}\gg
\dot a/a\sim\tau^{-1}$.  The large size of the Thomson drag terms in
the equations for $\dot\theta_\gamma$ and $\dot\theta_b$ of equations
(\ref{photon})--(\ref{baryon2}) make these equations numerically difficult
to solve.  Therefore, in this limit we shall follow the method of
Peebles \& Yu (1970) to obtain an alternative form of the equations
valid for $\tau_c/\tau\ll1$ and $k\tau_c\ll1$.  The starting point is
the exact equation obtained by combining the second of equations
(\ref{photon2}) and (\ref{baryon2}),
\begin{equation}
\label{momentum}
  (1+R)\dot\theta_b+{\dot a\over a}\theta_b-c_s^2 k^2\delta_b-
    k^2R\left(\frac{1}{4}\delta_\gamma-\Theta_\gamma\right)+
    R(\dot\theta_\gamma-\dot\theta_b)=(1+R)k^2\psi\ ,
\end{equation}
where the right-hand side is included only in the conformal Newtonian
gauge; in the synchronous gauge it is set to zero.  We have defined
$R\equiv(4/3)\bar\rho_\gamma/\bar\rho_b$.  We shall see that the terms
proportional to $(\dot\theta_\gamma-\dot\theta_b)$ and $\Theta_\gamma$
may be neglected to lowest order in $\hbox{max}\,[k\tau_c,\tau_c/\tau]$,
with the result that the baryons and photons behave like a single coupled
fluid with velocity $\vec v_b$.  However, we require a more accurate
approximation to account for the slip between the photon and baryon fluids.

 From the second of equations (\ref{photon2}), we have
\begin{equation}
\label{thetabdot}
  \theta_b-\theta_\gamma=\tau_c\left[\dot\theta_\gamma-k^2\left(
    \frac{1}{4}\delta_\gamma-\Theta_\gamma\right)-k^2\psi\right]
\end{equation}
in the conformal Newtonian gauge; in the synchronous gauge we simply set
$\psi=0$.  Writing $\dot\theta_\gamma$ as $\dot\theta_b+(\dot\theta_\gamma
-\dot\theta_b)$ and using equation (\ref{momentum}), we get
\begin{equation}
\label{slip}
  \theta_b-\theta_\gamma={\tau_c\over1+R}\left[-{\dot a\over a}
    \theta_b+k^2\left(c_s^2\delta_b-\frac{1}{4}\delta_\gamma+
    \Theta_\gamma\right)+\dot\theta_\gamma-\dot\theta_b\right]\ ,
\end{equation}
a result that is valid in both gauges.  From the third of equations
(\ref{photon}), we have
\begin{equation}
\label{f2-tca}
  \Theta_\gamma=\frac{\tau_c}{9}\left(\frac{8}{3}\theta_\gamma
    +\frac{4}{3}\dot h+8\dot\eta-10\dot\Theta_\gamma-\frac{3}{7}k
    F_{\gamma\,3}\right)
\end{equation}
in the synchronous gauge; in the conformal Newtonian gauge one sets
$\dot h=\dot\eta=0$.  We see that $\Theta_\gamma\sim\delta_\gamma
\times\hbox{max}\,[k\tau_c,\tau_c/\tau]$ (the case $k\tau_c$ corresponding
to acoustic oscillations with $\theta_\gamma\sim k\delta_\gamma$).
Higher moments of the photon distribution are smaller by additional
powers of $k\tau_c$ and we shall neglect them in the limit of tight
coupling considered here.  Our goal is to obtain equations for
$\dot\theta_b$ and $\dot\theta_\gamma$ that are accurate to second order
in $\tau_c$.

To get an equation for $\dot\theta_b$ we differentiate equation
(\ref{slip}) and use equation (\ref{momentum}).  Assuming that the gas
is nearly fully ionized so that $n_e\propto a^{-3}$ and that the baryon
temperature is approximately the radiation temperature implying $c_s^2
\propto a^{-1}$, we obtain
\begin{equation}
\label{sliprate}
  \dot\theta_b-\dot\theta_\gamma={2R\over1+R}{\dot a\over a}
    (\theta_b-\theta_\gamma)+{\tau_c\over1+R}\left[-{\ddot a\over a}
    \theta_b-{\dot a\over a}k^2\left(\frac{1}{2}\delta_\gamma+\psi\right)
    +k^2\left(c_s^2\dot\delta_b-\frac{1}{4}\dot\delta_\gamma\right)
    \right]+O(\tau_c^2) .
\end{equation}
This equation holds in the conformal Newtonian gauge; in the
synchronous gauge one should set $\psi=0$.  Note that $\dot\delta_b$
and $\dot\delta_\gamma$ are to be evaluated using the first of equations
(\ref{photon})--(\ref{baryon2}).  Substituting equation (\ref{sliprate})
into equation (\ref{momentum}) yields our desired equation of motion for
$\hbox{max}\,[k\tau_c,\tau_c/\tau]\ll1$.  If this condition is violated,
then one should use the explicit form of equations (\ref{baryon}) and
(\ref{baryon2}) for $\dot\theta_b$.

To obtain an equation for $\dot\theta_\gamma$ we combine the explicit
equations for $\dot\theta_\gamma$ and $\dot\theta_b$ to obtain the exact
equation
\begin{equation}
\label{thetagdot}
  \dot\theta_\gamma=-R^{-1}\left(\dot\theta_b+{\dot a\over a}\theta_b
    -c_s^2k^2\delta_b\right)+k^2\left(\frac{1}{4}\delta_\gamma-
    \Theta_\gamma\right)+{(1+R)\over R}k^2\psi
\end{equation}
in conformal Newtonian gauge; in synchronous gauge one sets $\psi=0$.
For $\dot\theta_b$ we use the tight-coupling approximation (substituting
eq. \ref{sliprate} into eq. \ref{momentum}) at early times and the exact
explicit equations (\ref{baryon}) or (\ref{baryon2}) at late times.
In practice, we switch to the explicit scheme for $\dot\theta_b$ when
$T_b=2\times10^4$ K; we switch to the explicit scheme for $\dot F_{\gamma
\,l}$ for $l>1$ and $T_b=2\times10^5$ K (at earlier times these moments
are set to zero).  We have verified that these switches occur early enough
to preserve good accuracy in the resulting photon phase space distribution.

\subsection{Recombination}

In order to compute the Thomson scattering terms in the equations of
motion for photons and baryons we need to know the free electron density
$n_e(\tau)$.  Our treatment is based on Peebles (1968; see also Peebles
1993).  We summarize the procedure here.

Because the Thomson opacity is enormous until hydrogen begins to recombine,
it is sufficient to treat the helium as being fully neutral at all times.
We define the ionization fraction of hydrogen as $x_e=n_e/n_H$ where $n_H$
is the number density of hydrogen nuclei.  The ionization rate equation
is (Peebles 1968; Spitzer 1978)
\begin{equation}
\label{ionrateeq}
  {d x_e\over d\tau}=aC_r\left[\beta(T_b)(1-x_e)-n_H\alpha^{(2)}(T_b)x_e^2
    \right]\ .
\end{equation}
The factor $C_r$ is discussed below.  The collisional ionization rate from
the ground state is
\begin{equation}
\label{ionrate}
  \beta(T_b)=\left(m_ek_{\rm B}T_b\over2\pi\hbar^2\right)^{3/2}
    e^{-B_1/k_{\rm B}T_b}\,\alpha^{(2)}(T_b)\ ,
\end{equation}
where $B_1=m_ee^4/(2\hbar^2)=13.6$ eV is the ground state binding energy,
and the recombination rate to excited states is
\begin{equation}
\label{recombrate}
  \alpha^{(2)}(T_b)={64\pi\over(27\pi)^{1/2}}{e^4\over m_e^2c^3}
    \left(k_{\rm B}T_b\over B_1\right)^{-1/2}\,\phi_2(T_b)\ ,\quad
    \phi_2(T_b)\approx0.448\,\ln\left(B_1\over k_{\rm B}T_b\right)\ .
\end{equation}
This expression for $\phi_2(T_b)$ is a good approximation (better
than one percent for $T_b<6000$ K).  At high temperatures this expression
underestimates $\phi_2$ but the neutral fraction is negligible
so that we make no significant error by setting $\phi_2=0$ for
$T_b>B_1/k_{\rm B}=1.58\times10^5$ K.

Recombination directly to the ground state is inhibited by the large
Lyman alpha and Lyman continuum opacities.  Net recombination must
occur either by 2-photon decay from the $2s$ level, with a rate
$\Lambda_{2s\to1s}=8.227\ {\rm s}^{-1}$, or by the cosmological redshifting
of Lyman alpha photons away from the line center.  Peebles (1968) gives
a detailed discussion of these atomic processes.  The net recombination
rate to the ground state is reduced by the fact that an atom in the
$n=2$ level may be ionized before it decays to the ground state.  The
reduction factor $C_r$ is just the ratio of the net decay rate (including
2-photon decay and Lyman alpha production at the rate allowed by redshifting
of photons out of the line) to the sum of the decay and ionization rates
from the $n=2$ level:
\begin{equation}
\label{cpeebles}
  C_r={\Lambda_\alpha+\Lambda_{2s\to1s}\over\Lambda_\alpha+\Lambda_{2s\to1s}
    +\beta^{(2)}(T_b)}\ ,
\end{equation}
where
\begin{equation}
\label{cpeebles2}
  \beta^{(2)}(T_b)=\beta(T_b)e^{+h\nu_\alpha/k_{\rm B}T_b}\ ,\quad
  \Lambda_\alpha={8\pi\dot a\over a^2\lambda_\alpha^3 n_{1s}}\ ,\quad
  \lambda_\alpha={8\pi\hbar c\over3B_1}=1.216\times10^{-5}\ {\rm cm}\ ,
\end{equation}
where $\nu_\alpha=c/\lambda_\alpha$.  For $T_b\ll 10^5$ K, it is a very
good approximation to replace the number density $n_{1s}$ of hydrogen
atoms in the $1s$ state by $(1-x_e)n_H$.

We integrate equation (\ref{ionrateeq}) using a stable and accurate
semi-implicit method with a large number of timesteps through recombination.
Since the results are independent of the perturbations, we pre-compute
the ionization history of a model and later use cubic splines interpolation
to obtain $x_e(\tau)$ accurately during integration of the perturbation
equations.

\section{Super-Horizon-Sized Perturbations and Initial Conditions}
\label{sec:gauge mode}

The evolution equations derived in the previous sections can be solved
numerically once the initial perturbations are specified.  We start
the integration at early times when a given $k$-mode is still outside
the horizon, i.e., $k\tau \ll 1$ where $k\tau$ is dimensionless.
(We follow common usage in referring to waves with $k\tau<1$ as being
``outside the horizon'' even though $\tau$ is more appropriately
called the comoving Hubble distance.)  The behavior of the density
fluctuations on scales larger than the horizon is gauge-dependent.
The fluctuations can appear as growing modes in one coordinate system
and as constant modes in another.  As we will show in this section,
this is exactly what occurs in the synchronous and the conformal
Newtonian gauges.

We first review the synchronous gauge behavior, which has already been
discussed by Press \& Vishniac (1980) and Wilson \& Silk (1981),
although these authors did not include neutrinos.  We are concerned
only with the radiation-dominated era since the numerical integration
for all the $k$-modes of interest will start in this era.  At this
early time, the massive neutrinos are relativistic, and the CDM
and the baryons make a negligible contribution to the total energy
density of the universe: $\bar{\rho}_{\rm total} = \bar{\rho}_\nu +
\bar{\rho}_{\gamma}$.  The expansion rate is $\dot{a}/a=\tau^{-1}$.
We can analytically extract the time-dependence of the metric and
density perturbations $h$, $\eta$, $\delta$, and $\theta$ on
super-horizon scales ($k\tau\ll1$) from equations (\ref{ein-syn}),
(\ref{massless}) and (\ref{photon}).  The large Thomson damping terms
in equations (\ref{photon}) drive the $l\ge 2$ moments of the photon
distribution function $F_{\gamma\,l}$ to zero.  Similarly, $F_{\nu\,l}$
for $l\ge3$ can be ignored because they are smaller than $F_{\nu\,2}$
by successive powers of $k\tau$.  Equations (\ref{ein-syna}),
(\ref{ein-sync}), (\ref{massless}), and (\ref{photon}) then give
\begin{eqnarray}
\label{press}
  && \tau^2\ddot{h} + \tau\dot{h} + 6 [(1-R_\nu)\delta_\gamma
	+R_\nu\delta_\nu] = 0 \,, \nonumber\\
  && \dot{\delta}_\gamma + {4\over 3}\theta_\gamma + {2\over 3}\dot{h}=0\,,
	\qquad \dot{\theta}_\gamma - {1\over 4}k^2 \delta_\gamma = 0
	\,,\nonumber\\
  && \dot{\delta}_\nu + {4\over 3}\theta_\nu + {2\over 3}\dot{h}=0\,,
   	\qquad \dot{\theta}_\nu - {1\over 4}k^2
	(\delta_\nu-4\Theta_\nu) = 0 \,,\\
  && \dot{\Theta}_\nu - {2\over 15}(2\theta_\nu+\dot{h}+6\dot{\eta}) = 0
	\,, \nonumber
\end{eqnarray}
where we have defined $R_\nu\equiv\bar\rho_\nu/(\bar\rho_\gamma
+\bar\rho_\nu)$.  For $N_\nu$ flavors of neutrinos ($N_\nu=3$ in the
standard model), after electron-positron pair annihilation and before
the massive neutrinos become nonrelativistic,
$\bar\rho_\nu/\bar\rho_\gamma=(7N_\nu/8)(4/11)^{4/3}$ is a constant.

To lowest order in $k\tau$, the terms $\propto k^2$ in equations
(\ref{press}) can be dropped, and we have $\dot{\theta}_\nu=\dot
{\theta}_\gamma =0$.   Then these equations can be combined into a
single fourth-order equation for $h$:
\begin{equation}
   \tau {d^4h\over d\tau^4} + 5{d^3h\over d\tau^3}=0\,,
\end{equation}
whose four solutions are power laws: $h \propto \tau^n$ with $n=0$, 1,
2, and $-2$.  From equations (\ref{press}) we also obtain
\begin{eqnarray}
\label{modes}
	h &=& A+B(k\tau)^{-2}+C(k\tau)^2 + D(k\tau) \,, \nonumber\\
	\delta\equiv(1-R_\nu)\delta_\gamma+R_\nu\delta_\nu &=&
		-{2\over 3}B(k\tau)^{-2} - {2\over 3}C(k\tau)^2
		- {1\over 6}D(k\tau) \,, \nonumber\\
	\theta\equiv(1-R_\nu)\theta_\gamma+R_\nu\theta_\nu &=&
		-{3\over 8}Dk \,,
\end{eqnarray}
and $A$, $B$, $C$, and $D$ are arbitrary dimensionless constants.
The other metric perturbation $\eta$ can be found from equation
(\ref{ein-syna}):
\begin{equation}
	\eta = 2C + {3\over 4}D(k\tau)^{-1} \,.
\end{equation}

Press \& Vishniac (1980) derived a general expression for the time
dependence of the four eigenmodes.  They showed that of these four
modes, the first two (proportional to $A$ and $B$) are gauge modes
that can be eliminated by a suitable coordinate transformation.  The
latter two modes (proportional to $C$ and $D$) correspond to physical
modes of density perturbations on scales larger than the Hubble distance
in the radiation-dominated era.  Both physical modes appear as growing
modes in the synchronous gauge, but the $C(k\tau)^2$ mode dominates at
later times.  In fact, the mode proportional to $D$ in the
radiation-dominated era decays in the matter-dominated era (Ratra 1988).
We choose our initial conditions so that only the fastest-growing
physical mode is present (this is appropriate for perturbations created
in the early universe), in which case $\theta_\gamma=\theta_\nu=
\dot\eta=0$ to lowest order in $k\tau$.  To get nonzero starting values
we must use the full equations (\ref{press}) to obtain higher
order terms for these variables.  To get the perturbations in the
baryons we impose the condition of constant entropy per baryon.
Using all of these inputs, we obtain the leading-order behavior
of super-horizon-sized perturbations in the synchronous gauge:
\medskip
\newline {\it Synchronous gauge ---\hfil}
\begin{eqnarray}
\label{super}
	&&\delta_\gamma = -{2\over 3}C (k\tau)^2 \,, \qquad
	  \delta_c = \delta_b = {3\over 4}\delta_\nu =
		{3\over 4}\delta_\gamma \,, \nonumber\\
	&&\theta_c=0\,,\quad
	\theta_\gamma = \theta_b = -{1\over 18}C(k^4 \tau^3)\,,
	\quad \theta_\nu={23+4R_\nu \over 15+4R_\nu}
	\theta_\gamma\,,\\
	&&\Theta_\nu= {4C \over 3(15+4R_\nu)} (k\tau)^2 \,,\nonumber\\
	&&h = C(k\tau)^2 \,,\qquad
	\eta = 2C - {5+4R_\nu\over 6(15+4R_\nu)}C(k\tau)^2\,.\nonumber
\end{eqnarray}
The initial conditions for the moments $\Psi_l, l \ge 1$, of the
massive neutrino distribution can be related to $\Psi_0$ and the
variables above by equations (\ref{massive}).  To obtain the initial
$\Psi_0$, we can write the perturbed distribution function as
$f=f_0(q)(1+\Psi_0)=2h_{\rm P}^{-3}\{\exp[q/k(T+\delta T)]+1\}^{-1}$,
where $\delta T/T=\delta_\nu/4$ by the isentropic condition.  Then using
equations (\ref{massive}), we find the first three moments to be
\begin{eqnarray}
\label{psil}
 \Psi_0 &=& - {1\over 4}\delta_\nu {d\ln f_0\over d\ln q}\,,\nonumber\\
 \Psi_1 &=& - {\epsilon\over qk}\theta_\nu {d\ln f_0\over d\ln q}\,,\\
 \Psi_2 &=& {5\over 2} \left[ {m_\nu^2 a^2\over 4q^2}\delta_\nu
   - {\epsilon^2\over q^2}\Theta_\nu\right] {d\ln f_0\over d\ln q}
	\,.\nonumber
\end{eqnarray}
The higher moments $\Psi_l$ ($l\ge 3$) are negligible for $k\tau\ll 1$.

The initial conditions for the isentropic perturbations in the conformal
Newtonian gauge can be obtained either by solving equations (\ref{ein-con}),
(\ref{massless2}), and (\ref{photon2}), or using the transformations given
by equations (\ref{trans2}) and (\ref{deltat}) (which enables us to relate
the amplitudes in the two gauges).  We find for the growing mode
\medskip
\newline {\it Conformal Newtonian gauge ---\hfil}
\begin{eqnarray}
\label{super2}
	&&\delta_\gamma = -{40 C\over 15+4R_\nu} = -2\psi \,,\qquad
	  \delta_c = \delta_b = {3\over 4}\delta_\nu =
		{3\over 4}\delta_\gamma \,, \nonumber\\
	&&\theta_\gamma=\theta_\nu=\theta_c=\theta_b={10 C\over
		15+4R_\nu} (k^2 \tau)=\frac{1}{2}(k^2\tau)\psi \,, \\
	&&\Theta_\nu= {4C \over 3(15+4R_\nu)} (k\tau)^2=\frac{1}{15}
		(k\tau)^2\psi \,,\nonumber\\
	&&\psi = {20 C\over 15+4R_\nu} \,,\qquad
	\phi = \left( 1+{2\over 5}R_\nu \right)\psi \,.\nonumber
\end{eqnarray}
The massive neutrino moments $\Psi_l$ in this gauge are related
to $\delta_\nu$, $\theta_\nu$, and $\Theta_\nu$ by the same equations
(eq.~\ref{psil}) as in the synchronous gauge.
As claimed earlier, $\psi=\phi$ to zeroth order in $k\tau$ when no
neutrinos are present (i.e., $R_\nu=0$).  If we characterize the
perturbations in the conformal Newtonian gauge by the potential
$\psi$, all matter and metric variables have a very simple form
outside the horizon.  The neutrino energy fraction $R_\nu$ enters only
in the second potential $\phi$ as a result of the shear stress produced
by the free-streaming neutrinos.  Bardeen (1980) was concerned that a
large shear stress would lead to large metric perturbations in the
conformal Newtonian gauge.  We see that this does not happen for
isentropic growing-mode perturbations in which the shear stress arises
solely due to the free-streaming of relativistic collisionless particles.

We see that $\delta$ grows with time in the synchronous gauge but
remains a constant in the conformal Newtonian gauge before horizon
crossing.  Another significant difference is the larger value of the
velocity perturbations for small $k\tau$ in the conformal Newtonian gauge.
Physically, this difference arises because velocity perturbations vanish
to lowest order in the synchronous gauge because the synchronous gauge
spatial coordinates are Lagrangian coordinates for freely-falling
observers (\S 2).  The next-order velocity perturbations differ for
the neutrinos and photons because these two fluids have effectively
different equations of state: the neutrinos are collisionless while
the photons behave like a perfect fluid due to their strong coupling
to the baryons.  In the conformal Newtonian gauge, the lowest-order
velocity perturbations do not vanish because the conformal Newtonian
gauge spatial coordinates are Eulerian coordinates.  If we were to
include the next-order corrections to $\theta$ proportional to
$k^4\tau^3$, differences between the different fluid components would
appear in equations (\ref{super2}).

In the conformal Newtonian gauge, the mode proportional to $D$ in the
synchronous gauge yields $\phi\propto\psi\propto\delta\propto(k\tau)^{-3}$.
Thus, this mode corresponds to a decaying mode in the conformal Newtonian
gauge even though it yields $\delta\propto(k\tau)$ in the synchronous gauge.
The two gauge modes ($A$ and $B$ in eq. \ref{modes}) do not exist in
the conformal Newtonian gauge.

\section{Integration Results in a CDM+HDM Model}
We apply the results derived in the previous sections to an $\Omega=1$
cosmological model consisting of a mixture of CDM and HDM, with
parameters $\Omega_{\rm cold}=0.65$, $\Omega_{\rm hot}= 0.3$,
$\Omega_{\rm baryon}=0.05$, and $H_0=50$ km s$^{-1}$ Mpc$^{-1}$.  The
corresponding neutrino mass is $m_\nu = 93.13\,(\Omega_{\rm hot} h^2)$
eV $= 6.985$ eV.

In Fourier space, all the $\vec k$ modes in the linearized Einstein,
Boltzmann, and fluid equations evolve independently; thus the equations
can be solved for one value of $\vec k$ at a time.  Moreover, all
modes with the same $k$ (the magnitude of the comoving wavevector)
evolve the same way.  We integrated the equations of motion numerically
over the range 0.01 Mpc$^{-1} \leq k \leq 100$ Mpc$^{-1}$ using 41
points evenly spaced in $\log_{10} k$ with an interval of $\Delta\log_{10}
k = 0.1$.  The time integration was performed using the standard fifth-
and sixth-order Runge-Kutta integrator {\tt dverk} (obtained from
netlib@research.att.com).  The time integration was begun at
conformal time $\tau_0= 3\times 10^{-4}$ Mpc with $z \sim 10^9$ and
ended at $\tau_f = 3000$ Mpc with $z = 13.55$.  The initial $\tau_0$
was chosen so that the largest $k$ (i.e. the smallest wavelength) was
well outside the horizon at the onset of the integration.  The
integration was stopped when the fluctuations were still in the
linear regime, with the rms density fluctuation in the CDM component
being about 0.2 (Ma \& Bertschinger 1994).

The Einstein equations provide redundant equations for the evolution
of the metric perturbations.  In the synchronous gauge we chose to use
$a\dot h$ and $\eta$ as the primary metric perturbation variables in
the integration, and used equation (\ref{ein-synb}) and a combination
of (\ref{ein-syna}) and (\ref{ein-sync}) as the evolution equations.
In the conformal Newtonian gauge we integrated $\phi$ using equation
(\ref{ein-conb}) and obtained $\psi$ algebraically using
(\ref{ein-cond}).  In both gauges we used the time-time Einstein
equation (eqs. \ref{ein-syna} and \ref{ein-cona}) to check integration
accuracy.  In the conformal Newtonian gauge it is possible to avoid
integration of the metric perturbations altogether by combining
equations (\ref{ein-cona}) and (\ref{ein-conb}) into an algebraic
equation for $\phi$.  However, we found that this gave numerical
difficulties because the initial value of $\phi$ has to be set with
exquisite precision when $k\tau\ll1$.  We also found it necessary to
obtain the initial $\theta$s from $\phi$ and the $\delta$s in the
combined constraint equations of (\ref{ein-cona}) and (\ref{ein-conb}).
Although the analytical expressions in equations (\ref{super2}) are
good approximations for $k\tau\ll 1$, slight deviations from the
energy-momentum constraints was found to cause numerical difficulties.

The photon and the massless neutrino phase space distributions were
expanded in Legendre series (see eq. \ref{fsubl}) with 1000 $l$-values
in order to guarantee sufficient angular resolution.  The massive
neutrinos are computationally expensive due to the momentum-dependence
in equations (\ref{massive}) and (\ref{massive2}).  We performed the
massive neutrino calculations on a grid of 100 $q$-points including 50
$l$-values for every $q$.  By setting the phase space harmonics to
zero for larger values of $l$, we found spurious waves which propagate
from low to high $l$, reflect at $l=50$, and then propagate back to
low $l$.  These numerical artifacts are analogous to the propagation
of traveling waves on a string with a fixed end.  However, only the
first three $l$-terms contribute directly to the source terms in the
Einstein equations, and we are not interested in the angular power
spectrum of the neutrinos themselves.  We found that $l_{\rm max}=50$
is adequate to ensure that the reflected waves from the cut-off at
$l_{\rm max}$ have not interfered with the low-$l$ harmonics.  We
checked that all of our numerical approximations are adequate by
increasing the grids of $k$, $q$, and $l$ values as well as decreasing
the integration timestep.  We estimate that our final results have a
relative accuracy better than $10^{-3}$.

We shall first present results for the metric perturbations in the
conformal Newtonian gauge and then compare the evolution of density
perturbations in our two gauges.  The metric perturbations in the
synchronous gauge have no simple physical interpretation, so we shall
not bother presenting them.

Figure 1 shows the time evolution of the metric
perturbations $\phi(k,\tau)$ and $\psi(k,\tau)$ in the conformal
Newtonian gauge for all 41 values of $k$.  The overall normalization
was chosen arbitrarily (corresponding to $C=-1/6$ in eqs. \ref{super}
and \ref{super2}).  The difference between $\psi$ and $\phi$ in the
radiation-dominated era is due to the shear stress contributed by the
relativistic neutrinos (eq. \ref{super2}) which make up a fraction
$R_\nu=0.4052$ of the total energy density.  On scales much smaller
than the horizon, $\psi$ corresponds to the Newtonian gravitational
potential and $\phi=\psi$ in the matter-dominated era.  As is well
known, the potential is constant for growing-mode density
perturbations of CDM in an Einstein-de Sitter universe.  In a mixed
dark matter model, however, the CDM density perturbation growth can be
suppressed by the lack of growth of HDM perturbations so that $\psi$
decays slowly.

The behavior of the metric perturbations can be understood as follows.
All of the 41 $k$-modes are outside the horizon at early times when
$\tau < 0.01$ Mpc.  The horizon eventually ``catches up'' and a given
$k$-mode crosses inside the horizon when $k\tau$ is about $\pi$.  The
modes with larger $k$ (i.e., shorter wavelengths) enter the horizon
earlier.  If a given $k$-mode enters the horizon during the
radiation-dominated era, the tight coupling between photons and
baryons due to Thomson scattering induces damped acoustic oscillations
in the conformal Newtonian gauge metric perturbations, which are
exhibited in Figure 1 by the modes with $k > 0.1$
Mpc$^{-1}$.  (In fact, it is not the speed-of-light horizon that sets
the scale for the oscillation and damping of the potential.  Rather,
as we show below, it is the acoustic horizon.  These horizons are
similar during the radiation-dominated era because the sound speed of
the photon-baryon fluid is $c/\sqrt{3}$.)  The modes with $k < 0.1$
Mpc$^{-1}$ enter the horizon during the matter-dominated era and do
not oscillate acoustically because the Jeans wavenumber $k_{\rm J}=
(4\pi G\bar\rho a^2/c_s^2)^{1/2}$ has then become much larger than the
wavenumbers under investigation.

We can understand the oscillations more quantitatively by studying the
Einstein equations (\ref{ein-con}) in the conformal Newtonian gauge.
Analytical solutions can be found for a perfect fluid with no shear
stress, in which case $\phi=\psi$.  Using $c_s^2 = \delta P/\delta\rho=
\bar p/\bar\rho$ and $\dot{a}/a =2\tau^{-1}/(1+3c_s^2)$ (from eqs.
\ref{friedmann} and \ref{friedmann2}), equations (\ref{ein-con}) can be
combined to yield
\begin{equation}
   \tau^2\ddot{\phi} + {6(1+c_s^2)\over
	1+3c_s^2}\tau\dot{\phi} + (kc_s\tau)^2\phi =0 \,,
\end{equation}
whose solutions are Bessel functions with a power-law pre-factor:
\begin{equation}
	\phi_\pm = (kc_s\tau)^{-\nu} J_{\pm\nu}(kc_s\tau)\,,
	\qquad \nu \equiv {5+3c_s^2 \over 2(1+3c_s^2)} \,.
\end{equation}
The $\phi_-$ solution corresponds to the decaying mode discussed in
\S 6, so we shall ignore it.  In the radiation-dominated era, $c_s^2
= \frac{1}{3}$ and $\nu=\frac{3}{2}$, so that
\begin{equation}
  \phi_+ = (kc_s\tau)^{-3/2} J_{3/2}(kc_s\tau)
	      \propto \cases{
		\mbox{constant}\,, & $kc_s\tau \ll 1$\,, \cr
		a^{-2} \cos(kc_s\tau)\,, & $kc_s\tau \gg 1$\,.\cr
		}
\end{equation}
This type of behavior is apparent in Figure 1 for $\tau <
\tau_{\rm eq}$.  This analytic solution holds of course only in the
absence of neutrinos; obtaining the correct amplitudes for $\psi$ and
$\phi$ in CDM+HDM models shown in Figure 1 required the full
integration discussed in this paper.

After the universe becomes matter-dominated, acoustic damping of the
potential ceases, and the only physical process causing the potential
to change is the free-streaming damping of perturbations in the massive
neutrinos.  We shall discuss this further after examining the evolution
of the density perturbations.

Figure 2 shows the evolution of the density perturbations for the five
particle species in the two gauges from our numerical integration.
Three wavenumbers are plotted: $k=0.01$ Mpc$^{-1}$ (Fig.~2a), $k=0.1$
Mpc$^{-1}$ (Fig.~2b), and $k=1.0$ Mpc$^{-1}$ (Fig.~2c).  Each mode is
normalized with the same initial amplitude for $\phi$ as in Figure 1.
There are several notable features:

\noindent {\it Before horizon crossing} ---

(1) The initial amplitudes of the $\delta$'s are related by the isentropic
initial conditions: $\delta_\gamma = \delta_\nu =\delta_h = 4
\delta_b/3 = 4 \delta_c /3$.  The behavior of $\delta$ outside the
horizon is strongly gauge-dependent.  In the synchronous gauge,
Figure 2 shows that all the $\delta$'s before horizon
crossing in the radiation-dominated era grow as $a^2$.  This confirms
equations (\ref{super}) since $a(\tau)\propto \tau$ at this time.
It is straightforward to show that in the matter-dominated era (for
$\Omega$=1), $\delta \propto \tau^2 \propto a$ for all modes before
horizon crossing.  In the conformal Newtonian gauge, the $\delta$'s
remain constant outside the horizon as derived in equations (\ref{super2}).
After horizon crossing, however, the perturbations come into causal
contact and become nearly independent of the coordinate choices.
As one can see, $\delta_c$, $\delta_b$, and $\delta_h$ in the two gauges
are almost identical at late times.  In fact, one can show using
equations (\ref{ein-con}) and (\ref{deltat}a) that if $\psi=\hbox
{constant}$, then $\delta(\mbox{Syn})-\delta(\mbox{Con})=2\psi$,
accounting for the slight differences apparent in the figures.

\noindent {\it After horizon crossing} ---

(2) For CDM, the $k$-modes that enter the horizon during the
radiation-dominated era behave very differently from those entering in
the matter-dominated era.  The critical scale separating the two is
the horizon distance at the epoch of radiation-matter equality
($a_{\rm eq} \sim 10^{-4}$): $k_{\rm eq} = 2\pi/\tau_{\rm eq} \sim
0.1$ Mpc$^{-1}$ for our parameters.  For the modes with $k > 0.1$,
horizon crossing occurs when the energy density of the universe is
dominated by radiation; thus the fluctuations in the CDM can not grow
appreciably during this time.  For the photons and the baryons, the
important scale is the horizon size at recombination ($a_{\rm rec}
\sim 10^{-3}$): $k_{\rm rec} = 2\pi/\tau_{\rm rec}
\sim 0.025$ Mpc$^{-1}$.  The modes with $k > 0.025$ (see
Figs.~2b and 2c) enter the horizon
before recombination, so the photons (long-dashed curves) and
baryons (dash-dotted curves) oscillate acoustically while they are
coupled by Thomson scattering.  The coupling is not perfect.
The friction of the photons dragging against the baryons leads
to Silk damping (Silk 1968), which is prominent in Figure 2c
at $a\sim10^{-3.5}$.  The baryons decouple from the photons at
recombination and then fall very quickly into the potential wells
formed around the CDM, resulting in the rapid growth of $\delta_b$
in Figures 2b and 2c.

(3) Neutrinos decouple from other species at $T \sim 1$ MeV and $a
\sim 10^{-10}$.  At this early time, both the massless and the 7 eV
massive neutrinos behave like relativistic collisionless particles.
The massive neutrinos become non-relativistic when $3k_{\rm B}T_\nu \sim
m_\nu$, corresponding to $a_{\rm nr} \sim 7\times 10^{-5}$ for
7 eV neutrinos.  Close inspection of the figures at $a\approx
a_{\rm nr}$ reveals that $\delta_h$ (short-dashed curve) is
indeed making a gradual transition from the upper line for the
radiation fields to the lower line for the matter fields.  Although
the Jeans length of a fluid is not well defined for collisionless
particles such as the neutrinos, the criterion for free-streaming
damping is similar to the Jeans criterion for gravitational stability:
free-streaming is important for $k > k_{\rm fs}$, where $k^2_{\rm fs}(a)
=4\pi G\bar\rho a^2/\langle v^2\rangle$ and $\langle v^2\rangle^{1/2}$
is the neutrino velocity dispersion.  When the neutrinos are relativistic,
$v \sim 1$ and $k_{\rm fs}(a) \propto a^{-1}$ for $a < a_{\rm eq}$.
After the neutrinos become non-relativistic, the neutrino speed is
given by $\langle v^2\rangle^{1/2} \sim 3 k_{\rm B} T_\nu/m_\nu =
3 k_{\rm B} T_{0,\nu}/a\,m_\nu\,$, implying $\langle v^2\rangle^{1/2}
\sim 15 a^{-1} (m_\nu/10\,{\rm eV})^{-1}$ km s$^{-1}$.  In the
matter-dominated era, we have $4\pi G\bar\rho a^2 = \frac{3}{2} H_0^2
a^{-1}$ from the Friedmann equation, and therefore
\begin{equation}
\label{kfs}
	k_{\rm fs}(a) = 8\,a^{1/2}\,\left({m_\nu \over 10
	{\rm eV}}\right) h\,{\rm Mpc}^{-1}  \,.
\end{equation}
In Figure 2a, since horizon crossing occurs when the
free-streaming effect is already unimportant, the evolution of
$\delta_h$ is very similar to that of CDM.  In Figures 2b
and 2c, however, the free streaming effect is evident
and the growth of $\delta_h$ is suppressed until $k_{\rm fs}(a)$ grows
to $\sim k$. After $k_{\rm fs}(a) > k$, $\delta_h$ can grow again and
catch up to $\delta_c$.  Since $k_{\rm fs} \propto a^{1/2}$, the larger
$k$ modes suffer more free-streaming damping and $\delta_h$ can not
grow until later times.  The damping in $\delta_h$ also affects the
growth of $\delta_c$, slowing it down more for larger $\Omega_{\rm hot}$
compared to the pure CDM model.  This effect is also apparent in the power
spectra shown in, for example, Ma \& Bertschinger (1994).

\section{Summary}
Physical quantities are independent of the coordinate systems they are
computed in.  For historical reasons, most calculations of linear
fluctuation growth have been carried out in the synchronous gauge.  In
this paper we explored an alternative gauge, the conformal Newtonian
gauge, which is free of the gauge ambiguities and coordinate
singularities associated with the synchronous gauge.  We derived the
coordinate transformation relating the two gauges and presented the
linear theory of isentropic scalar gravitational perturbations in
parallel for both gauges.  The complete set of evolution equations are
given: the Einstein equations for the metric perturbations, the
Boltzmann equations for the photon and neutrino phase space
distributions, and the fluid equations for CDM and baryons.

The use of the conformal Newtonian gauge was motivated by our work on
the HDM initial conditions in CDM+HDM models (Ma \& Bertschinger
1994).  In order to sample the neutrino phase space accurately for an
N-body simulation, we followed individual neutrino trajectories by
numerically integrating the geodesic equations in a perturbed
Friedmann-Robertson-Walker background metric.  The conformal Newtonian
gauge proved to be the most convenient choice for this calculation
because the geodesic equations have simple forms and the coordinates
do not become seriously deformed at late times.

In this paper we applied the linear theory to the CDM+HDM model under
study: $\Omega_{\rm cold}=0.65$, $\Omega_{\rm hot}=0.3$, $\Omega_{\rm
baryon}=0.05$, and $H_0$=50 km s$^{-1}$ Mpc$^{-1}$.  The evolution of
the density fields for all five particle species in both gauges was
presented.  We also illustrated the gauge dependence of the density
fields before horizon crossing and discussed the physical
interpretation of the results.

Interested users may obtain our programs to integrate the perturbation
equations by contacting one of the authors.

\acknowledgments

We appreciate the advice and comments of Alan Guth.  This work was
supported by NSF grant AST90-01762 and DOE grant DE-AC02-76ER03069.
The numerical work was carried out in part on the Convex C3880
provided by the National Center for Supercomputing Applications.
C.-P. Ma acknowledges Fellowship support from the Division of Physics,
Mathematics, and Astronomy at Caltech.  E.B. would like to thank John
Bahcall for his hospitality at the Institute for Advanced Study during
the completion of this work.

\clearpage

%\section{Figure Captions}
%
%\noindent Fig.~\ref{fig:phi}:
%The scalar metric perturbations $\phi(k,\tau)$ (Fig.~\ref{fig:phi}a) and
%$\psi(k,\tau)$ (Fig.~\ref{fig:phi}b) in the conformal
%Newtonian gauge as a function of $\tau$.  The 41 curves
%from left to right correspond to 41 values of $k$
%between 100.0 Mpc$^{-1}$ and 0.01 Mpc$^{-1}$.  The labels
%$\tau_{\rm nr}$, $\tau_{\rm eq}$ and $\tau_{\rm rec}$ indicate
%the time 7 eV neutrinos become non-relativistic, the matter-radiation
%equality time, and the recombination time, respectively.
%
%\noindent Fig.~\ref{fig:bolt}:
%Evolution of the density fields in the synchronous gauge (top panels)
%and the conformal Newtonian gauge (bottom panels) for 3 wavenumbers:
%$k$= 0.01 (Fig.~\ref{fig:bolt}a), 0.1 (Fig.~\ref{fig:bolt}b), and 1.0
%(Fig.~\ref{fig:bolt}c) Mpc$^{-1}$.  In each figure, the five lines
%represent $\delta_c$, $\delta_b$, $\delta_\gamma$, $\delta_\nu$, and
%$\delta_h$ for the CDM (solid curve), baryon (dash-dotted), photon
%(long-dashed), massless neutrino (dotted), and massive neutrino
%(short-dashed) components, respectively.
%
%\clearpage
%
%%%% Insert Figure phi
%\begin{figure}[p]
\begin{figure}
   \label{fig:phi}
   \vspace{4.5truein}
%temp \special{psfile=phi.ps voffset=-210 hoffset=-65 vscale=0.85 hscale=0.85}
   \includegraphics{phi.ps}
\end{figure}
%%%
\begin{figure}
   \vspace{4.5truein}
   \caption{The scalar metric perturbations $\phi(k,\tau)$ (Fig.~1a) and
	$\psi(k,\tau)$ (Fig.~1b) in the conformal Newtonian gauge as a
	function of $\tau$.  The 41 curves from left to right correspond
	to 41 values of $k$ between 100.0 Mpc$^{-1}$ and 0.01 Mpc$^{-1}$.
	The labels $\tau_{\rm nr}$, $\tau_{\rm eq}$ and $\tau_{\rm rec}$
	indicate the time 7 eV neutrinos become non-relativistic, the
	matter-radiation equality time, and the recombination time,
	respectively.}
%temp \special{psfile=psi.ps voffset=-270 hoffset=-65 vscale=0.85 hscale=0.85}
   \includegraphics{psi.ps}
\end{figure}
%%%%%%
%%%% Insert Figure bolt.k.ps
\begin{figure}
   \label{fig:bolt}
%temp   \vspace{4.truein}
%temp   \special{psfile=syn.1b.ps voffset=-270 hoffset=-65 vscale=0.85
%temp	hscale=0.85}
%temp   \vspace{4.5truein}
%temp   \special{psfile=consyn.1b.ps voffset=-300 hoffset=-65 vscale=0.85
%temp	hscale=0.85}
  \vspace{3.1truein}
  \includegraphics{syn.1b.ps}
  \vspace{4.2truein}
  \includegraphics{consyn.1b.ps}
%  \vspace{3.5truein}
%  \caption{Figure 2a}
\end{figure}
%%%%%%%%%%%%
\begin{figure}
%temp   \vspace{4.truein}
%temp   \special{psfile=syn.11b.ps voffset=-320 hoffset=-65 vscale=.85
%temp	hscale=.85}
%temp   \vspace{4.5truein}
%temp   \special{psfile=consyn.11b.ps voffset=-320 hoffset=-65 vscale=.85
%temp	hscale=.85}
   \vspace{3.1truein}
    \includegraphics{syn.11b.ps}
   \vspace{4.2truein}
   \includegraphics{consyn.11b.ps}
\end{figure}
%%%%%%%%%%%%%
\begin{figure}
%temp   \vspace{4.truein}
%temp   \special{psfile=syn.21b.ps voffset=-320 hoffset=-65 vscale=.85
%temp	hscale=.85}
%temp   \vspace{4.5truein}
%temp   \special{psfile=consyn.21b.ps voffset=-320 hoffset=-65 vscale=.85
%temp	hscale=.85}
   \vspace{3.truein}
   \includegraphics{syn2000.21b.ps}
   \vspace{4.03truein}
   \includegraphics{con2000.21b.ps}
   \vspace{.7truein}
   \caption{Evolution of the density fields in the synchronous
	gauge (top panels) and the conformal Newtonian gauge (bottom
	panels) for 3 wavenumbers $k$= 0.01 (Fig.~2a), 0.1 (Fig.~2b) and
	1.0 (Fig.~2c) Mpc$^{-1}$.  In each figure, the five lines represent
	$\delta_c, \delta_b, \delta_\gamma, \delta_\nu$ and $\delta_h$
	for the CDM (solid curve), baryon (dash-dotted), photon
	(long-dashed), massless neutrino (dotted), and massive neutrino
	(short-dashed) components, respectively.}
\end{figure}
%%%%%%

\end{document}